\documentclass[12pt]{iopart}
\usepackage{amsfonts}
\usepackage{graphicx}

\usepackage{color}
\usepackage{subfigure}
\usepackage{textcomp}
\usepackage{cite}
\usepackage{float}
\usepackage{soul}
\usepackage{url}

\definecolor{dgreen}{rgb}{0,0.7,0}

\expandafter\let\csname equation*\endcsname\relax
\expandafter\let\csname endequation*\endcsname\relax
\usepackage{amsmath,amssymb}

\usepackage{esint}

\newcommand{\ee}{\mathrm{e}}

\newcommand{\al}{\alpha}
\newcommand{\ep}{\epsilon}

\begin{document}
\title[Extreme statistics and index distribution in the classical $1d$ Coulomb gas]{Extreme statistics and index distribution in the classical $1d$ Coulomb gas}

\author{Abhishek Dhar}
\address{International Centre for Theoretical Sciences, TIFR, Bangalore 560089, India}

\author{Anupam Kundu}
\address{International Centre for Theoretical Sciences, TIFR, Bangalore 560089, India}

\author{ Satya N. Majumdar}
\address{LPTMS, CNRS, Univ. Paris-Sud, Universit\'{e} Paris-Saclay, 91405 Orsay, France}

\author{Sanjib Sabhapandit}
\address{Raman Research Institute, Bangalore 560080, India}

\author{Gr\'egory Schehr}
\address{LPTMS, CNRS, Univ. Paris-Sud, Universit\'{e} Paris-Saclay, 91405 Orsay, France}

\date{\today}


\begin{abstract}
%
We consider a one-dimensional gas of $N$ charged particles confined by an external harmonic potential and interacting via the one-dimensional Coulomb potential. For this system we show that in equilibrium the charges settle, on an average, uniformly and symmetrically on a finite region centred around the origin. We study the statistics of the position of the rightmost particle $x_{\max}$ and show that the limiting distribution describing its typical fluctuations is different from the Tracy-Widom distribution found in the one-dimensional log-gas. We also compute the large deviation functions which characterise the atypical fluctuations of $x_{\max}$ far away from its mean value. In addition, we study the gap between the two rightmost particles as well as the index $N_+$, i.e., the number of particles on the positive semi-axis. We compute the limiting distributions associated to the typical fluctuations of these observables as well as the corresponding large deviation functions. We provide numerical supports to our analytical predictions. Part of these results were announced in a recent Letter, {\it Phys. Rev. Lett.} {\bf 119}, 060601 (2017).
\end{abstract}



\section{Introduction}
In the last two decades, extreme value statistics in correlated random variables has received a resurgence of interest \cite{Buchanan14, Wolchover14} with the discovery of the Tracy-Widom (TW) distribution in the context of random matrix theory (RMT) \cite{Tracy94, Tracy96}. Since then the TW distribution has appeared ubiquitously in physics \cite{Majumdar07, Kriecherbauer10}, mathematics \cite{Baik99, Johnstone01} and information theory \cite{Kazakopoulos11}.  In physics it has appeared in stochastic growth models belonging to the Kardar-Parisi-Zhang (KPZ) universality class \cite{Johansson00, Prahofer00, Majumdar04,Sasamoto10,Calabrese10,Dotsenko10,Amir11}, nonintersecting Brownian motions \cite{Forrester11}, noninteracting fermions in a one-dimensional trapping potential \cite{Eisler13,Dean15,Dean16}, disordered mesoscopic systems \cite{Vavilov01} and even in the Yang-Mills gauge theory in two dimensions \cite{Forrester11}. It has also been measured experimentally in several systems including liquid crystals \cite{Takeuchi11}, coupled fiber lasers \cite{Fridman12}, or disordered superconductors \cite{Lemarie13}. 

Originally, the TW distribution was discovered as the limiting distribution of the largest eigenvalue $x_{\max}$ of an $N \times N$ Gaussian random matrix for which the joint probability density function (PDF) $\mathcal{P}(\{x_i\})$ of the $N$ real eigenvalues $\{x_1,x_2,...,x_N \}$ is known explicitly  \cite{Mehta91, Forrester10}: 
\begin{equation}
\mathcal{P}(\{x_i\}) = B_N \, \e^{-\frac{1}{2\sigma^2} \sum_{i=1}^N x_i^2} \prod_{i<j} |x_i - x_j|^\beta
\label{JPDF-rm}
\end{equation}
where $B_N$ is the normalisation constant and $\beta=1,2,4$ is the Dyson index corresponding respectively to the Gaussian Orthogonal, Unitary and Symplectic ensembles (GOE, GUE, and GSE respectively)~\cite{Mehta91, Dyson62}. This distribution of $N$ eigenvalues can, equivalently, be interpreted as the equilibrium Gibbs distribution, $\mathcal{P}(\{x_i\})=B_N~e^{-\beta E(\{x_i\})}$, of a gas of charged particles with positions $x_i$'s on a line with the energy $E(\{x_i\})$ given by
\begin{equation}
E=  \frac{1}{2\sigma^2 \beta} \sum_{i=1}^N x_i^2 - \frac{1}{2}\sum_{i \neq j}\log (|x_i-x_j|) \;. \label{H-log-gas-unscaled}
\end{equation} 
The first term in the energy can be interpreted as the potential energy due to a confining harmonic potential, while the second term represents a logarithmic repulsion between any pair of charges. These two opposite energies compete with each other. The first term scales for large $N$ as $\sim N\, x_{\rm typ}^2/(\sigma^2 \beta)$ where $x_{\rm typ}$ is the typical scale of the position of charges. The second term scales as $N^2$ since there are $N(N-1)$ pair of charges. Balancing the two energies lead to the fact that $x_{\rm typ} \sim \sqrt{N}$ for large $N$. This suggests a rescaling of the positions of the charges as $x_i \to \sigma \sqrt{\beta N}\, x_i$. In these rescaled variables, the energy is then given by
\begin{equation}
E=  \frac{1}{2} \left[N\,\sum_{i=1}^N x_i^2 - \sum_{i \neq i} \log |x_i - x_j| \right]\;, \label{H-log-gas}
\end{equation} 
up to an unimportant constant. This system is often known as the log-gas~\cite{Forrester10}. In the large $N$ limit, the average density $\rho_N(x)$ of these charges or the eigenvalues converges to an $N$-independent limiting density given by the Wigner semi-circular form
\begin{equation}
\rho_N(x)|_{N \to \infty}=\rho_\infty(x) = \frac{\sqrt{2-x^2}}{\pi}, \label{wig-sc}
\end{equation} 
which has a finite support $x \in [-\sqrt{2},~\sqrt{2}]$. It turns out that the behaviour of the eigenvalues close
to the soft edges $\rm \sqrt{2}$ have universal features -- most easily demonstrated by the largest eigenvalue $x_{\max} = \max_{1 \leq i \leq N} x_i$. In the log-gas picture, $x_{\max}$ corresponds to the position of the rightmost charge. Its average value is $\langle x_{\max} \rangle \sim \sqrt{2}$ for large $N$ and it corresponds to the right edge of the semi-circle. However, $x_{\max}$ typically fluctuates from sample to sample on a scale of width $N^{-2/3}$ around the mean. The probability distribution of these typical fluctuations is described by the celebrated TW distribution. Indeed, the cumulative distribution $Q(w,N) = {\rm Prob}(x_{\max}\leq w,N)$, takes the scaling form for $w - \sqrt{2} = O(N^{-2/3})$ 
\begin{eqnarray}
Q(w,N) \approx {\cal F}_\beta \left(\sqrt{2}N^{2/3}(w-\sqrt{2})\right) \label{TW} \;,
\end{eqnarray}
where ${\cal F}_\beta(x)$ is the TW distribution, computed by Tracy and Widom for $\beta = 1, 2$ and $4$
in terms of the solution of a Painlev\'e II equation~\cite{Tracy94}. For example, for $\beta = 2$ (the GUE case) 
\begin{eqnarray}
\mathcal{F}_2(x)&=& \exp \left(- \int_x^\infty (y-x)q^2(y)~dy \right),~~~\text{where},\nonumber \\
\frac{d^2q(y)}{dy^2}&=&2 q(y)^3 + y~q(y),~~\text{with},~~q(y \to \infty) \to {\rm Ai}(y), \label{TW-2}
\end{eqnarray}
 and ${\rm Ai}(y)$ is the Airy function. For general $\beta$, the PDF ${\cal F}'_\beta(x)$ of the TW-scaling function has non-Gaussian tails 
\begin{equation}
\mathcal{F}'_\beta(x) \sim
\begin{cases}
&\exp\left[-\frac{\beta}{24}|x|^3 \right]~~\text{as}~~x \to - \infty \\
& \\
&\exp\left[-\frac{2\beta}{3}x^{3/2} \right]~~\text{as}~~x \to  \infty.
\end{cases}
\label{TW-asymptotic}
\end{equation}
While the typical fluctuations of $x_{\max}$ around its mean are described by the TW distribution, the {\it atypical} large fluctuations of $x_{\max}$, far from its mean to the left and right, are not described by TW but rather by the left and right large deviation tails 
\begin{equation}\label{regimes_gaussian_CDF}
\partial_w\,Q(w,N)\approx
\begin{cases}
\exp\left[-\beta N^2 \Phi_{-}\left(w\right)\right]&,\,  
w<\sqrt{2} \; \& \; |w-\sqrt{2}|\sim\mathcal{O}(1)\\
\\
\sqrt{2} 
N^{\frac{2}{3}}{\cal F}'_\beta \left(\sqrt{2} 
N^{\frac{2}{3}}(w-\sqrt{2})\right)&, \, 
\hspace*{2.1cm} |w-\sqrt{2}|\sim\mathcal{O}(N^{-\frac{2}{3}})\\
\\
\exp\left[-\beta N \Phi_{+}\left(w\right)\right]&, 
\, w>\sqrt{2} \, \& \, |w-\sqrt{2}|\sim\mathcal{O}(1) \;,
\end{cases}
\end{equation}
where the right large deviation function (LDF)  $\Phi_+(w)$ was obtained explicitly for $\beta = 1$ in \cite{Arous01} and for arbitrary $\beta$ in \cite{Majumdar09}. On the other hand, $\Phi_-(w)$ was computed explicitly for all $\beta$ in \cite{Dean06,Dean08}. It was argued that the left and the right large deviation tails can be interpreted as the free energies of two different thermodynamic phases of the Coulomb gas, separated by a third order phase transition \cite{Majumdar14} in the large $N$ limit. Similar third order phase transitions have also been found in a variety of other systems \cite{Nadal11,Majumdar14, Schehr13, Colomo13,Ledoussal16,Sasorov17}, including in higher dimensions $d\geq 1$ \cite{Allez14, Cunden16, Cunden17a,Cunden17b}.  
 
The TW distribution was initially derived for an harmonic potential. However, it was found later that the typical distribution of $x_{\max}$ is universally given by the TW distribution, irrespective of the shape of the confining potential. This holds provided the average charge/eigenvalue density has a finite support and moreover, vanishes as a square root at the upper edge of the support (for a recent review see \cite{Bourgade}). One then naturally asks the question: what happens to the universality of the TW distribution, if instead of the confining potential, one changes the form of the repulsive pairwise interaction? A natural setting to address this question corresponds to a model in $1d$ of $N$ charged particles in presence of a confining harmonic potential and just replacing the logarithmic pairwise repulsion by the true Coulomb repulsion in $1d$, i.e. a linear $|x_i - x_j|$ interaction term in Eq. (\ref{H-log-gas}) instead of $\log |x_i - x_j|$. This well known model of $1d$ Coulomb gas has been studied earlier in the context of 1d charged plasma \cite{Choquard1981}. It is known as the one dimensional one component plasma ($1d$ OCP) or the ``jellium'' model, where $N$ charges of the same sign interact in the presence of a uniform background of opposite charges, assuring charge neutrality.   This model is a paradigm for $1d$ charged plasma \cite{Choquard1981} as several observables can be  calculated analytically~\cite{ Lenard61, Prager62, Baxter63, Dean10,TT2015}.  For this model most of the earlier studies considered bulk properties at the thermodynamic limit. In a recent Letter \cite{Dhar17}, we addressed the extreme value question in the $1d$ OCP or the ``jellium'' model where we showed analytically that the limiting distribution of the typical fluctuations of $x_{\max}$ is indeed {\it different} from the TW distribution. Moreover, by computing the left and the right LDFs explicitly, we have shown that the third-order phase transition between a pushed gas (left large deviation) and a pulled gas (right large deviation) is still present in this system as in the case of the log-gas. One of the purposes of the current paper is to provide a detailed derivation of these results presented in the Letter~\cite{Dhar17}.

In fact, the question of universality with respect to the pairwise repulsion term is not restricted just to the rightmost particle position $x_{\max}$, but can also be addressed for other observables. For instance, one can ask how sensitive is the statistics of the gap $g$ between the positions of the rightmost and the next rightmost particles for large $N$, as one changes the form of the pairwise interaction? Indeed, for the GUE, the PDF of the gap takes the scaling form, for large $N$ 
\begin{align}
P_G(g,N) \approx \sqrt{2}N^{2/3}~h_2(g \sqrt{2}N^{2/3}) \;, \label{gap_GUE}
\end{align} 
where the scaling function $h_2(x)$ was computed explicitly in \cite{Bornemann_Forrester_Whitte, Perret14}. In this paper, we compute exactly the gap distribution in the ``jellium'' model and show that it is different from the GUE-log-gas in Eq. (\ref{gap_GUE}). 

Another interesting observable that has been studied extensively in the context of random matrices is the index $N_+$, that denotes the number of positive eigenvalues, or equivalently the number of charges on the positive semi-axis. Obviously $N_+$ is a random variable with values $0 \leq N_+ \leq N$. It was shown that $N_+$ typically fluctuates around its mean value $N/2$ on a scale of width $\sqrt{\log N}$ and the typical fluctuations are given by a Gaussian form \cite{Cavagna00,Majumdar09b,Majumdar11}
\begin{align}
P_I(N_+,N) \approx \exp \left[ - \frac{\beta \pi^2}{2 \ln N}(N_+-N/2)^2\right] \;. \label{gauss_index}
\end{align} 
The atypical large deviations of $(N_+ - N/2) = O(N)$ were computed for large $N$ and it was found that \cite{Majumdar09b,Majumdar11}
\begin{align}\label{index_ldf}
P_I(N_+,N) \sim \e^{- \beta N^2 \Psi\left( \frac{N_+}{N }\right)  }
\end{align}  
where the rate function $\Psi(c)$ has a logarithmic singularity at $c=1/2$. In this paper, we study this index distribution analytically for the ``jellium'' model and find that it is rather different from the log-gas case. 
 
Thus the purpose of this paper is essentially twofold:
\begin{itemize}
\item[$\bullet$] to present the detailed calculations of the distribution of $x_{\max}$ in the ``jellium'' model
\item[$\bullet$] to present new exact results for the distribution of two other observables in the ``jellium'' model: (i) the gap $g$ between the positions of the two rightmost particles and (ii) the index $N_+$ denoting the number of particles on the positive semi-axis.  
\end{itemize}
Interestingly, as we will show, the function that characterises the limiting distribution of $x_{\max}$ happens to 
also characterise the limiting distribution of the gap and that of the index. One of the main results of this paper is to show that changing the pairwise interaction indeed changes significantly the behaviour of these observables, thereby changing their universality class.

\section{Model definition and the summary of the results}
\label{Summary}

We consider $N$ charges on a line, with positions $\{ x_i\}$, confined by an external harmonic potential and interacting pairwise via the true $1d$ repulsive Coulomb potential. We assume that the system is in thermal equilibrium such that the probability to observe the system in a configuration $\{ x_i\}$ is given by the Boltzmann distribution 
\begin{eqnarray}
\mathcal{P}(x_1,x_2,...,x_N) = \frac{\exp \left[-\beta\,E(x_1,x_2,...,x_N)\right]}{Z_N}, \label{jpdf-1dc}
\end{eqnarray}
where $\beta$ is the inverse temperature, $Z_N$ is the normalization constant and the energy of the configuration is given by
\begin{equation}
E(\{x_i\})=\frac{N}{2} \sum_{i=1}^N x_i^2 - \alpha  \sum_{i \neq j} |x_i-x_j| \;, \label{E-1dc}
\end{equation}
where $\alpha \geq 0$ denotes the strength of the Coulomb repulsion. As in the log-gas case in Eq. (\ref{H-log-gas}), the prefactor $N$ in the first term ensures that the $x_i$'s are of order $O(1)$. 

\begin{figure}[t]
  \centering
      \includegraphics[scale=0.4]{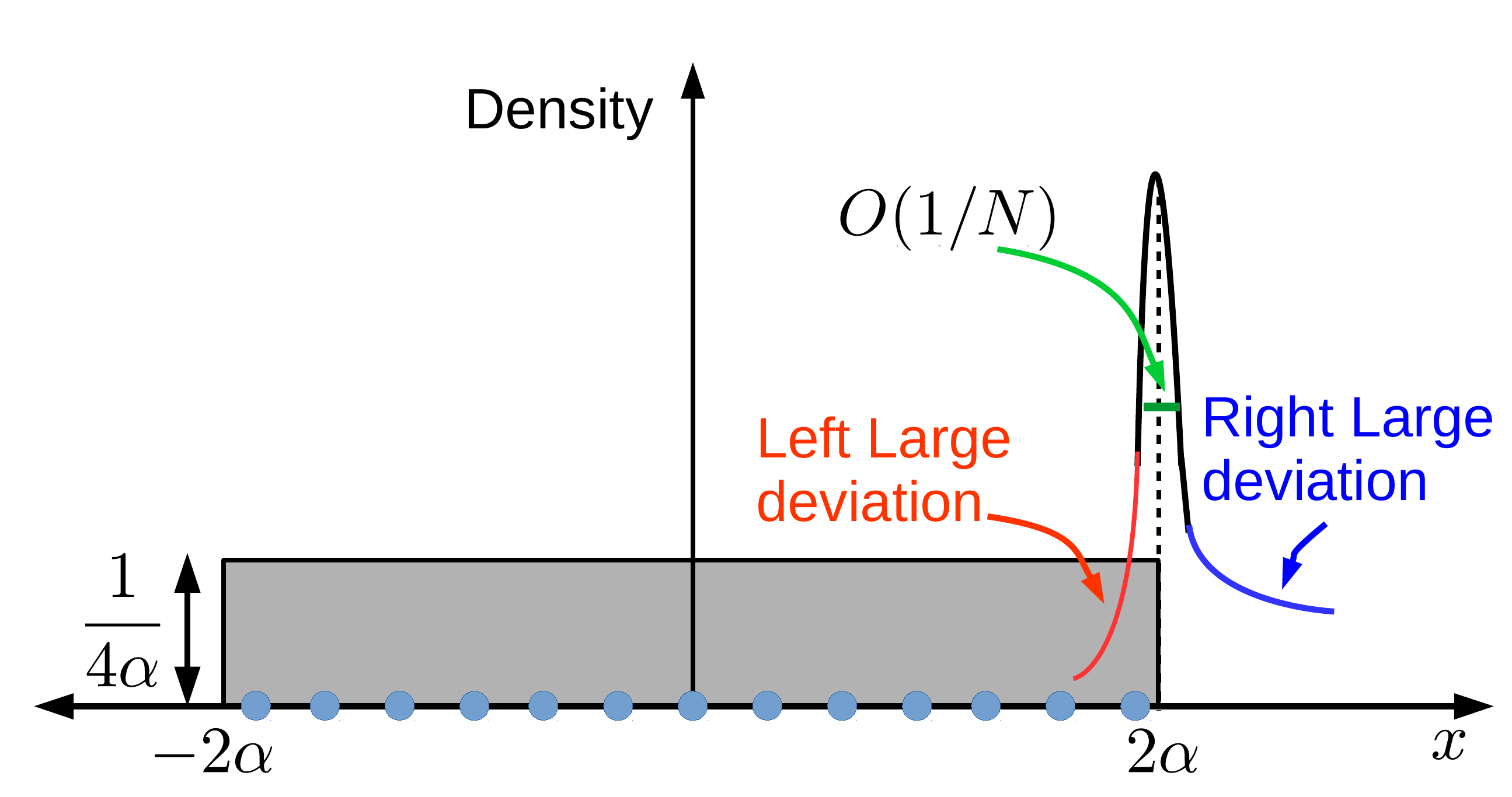}
  \caption{Schematic plot of the flat average density profile and the PDF of $x_{\max}$ in the thermodynamic limit. The PDF is peaked around the right edge $2\al$ of the average density profile. The position $x_{\max}$ fluctuates typically around the mean $2\alpha$ over the scale ${O}(1/N)$ for $\beta_0 \sim O(N)$ and these fluctuations are described by $F_\alpha'(x)$ [see \eqref{limit.1}], while the large deviations of ${O}(1)$ to the left and right of the mean are described by the left (red) and right (blue) large deviation tails.}
\label{fig1}
\end{figure}

Let us first consider two limiting temperature regimes: (i) very high temperature when $\beta \ll N$ and (ii) very low temperature with $\beta \gg N$. In the former case, the interaction between the charges become totally irrelevant and the particles behave as $N$ independent random variables with Gaussian distributions. In contrast, in case (ii) the interaction term dominates and the positions of the particles get ``frozen'' at equidistant points in the interval $[-2\alpha, + 2 \alpha]$, with very small fluctuations around them. It turns out that the most interesting situation occurs 
when $\beta = O(N)$ when both the interaction term as well as the confining potential compete with each other. Henceforth, in this paper, we will focus on the case where $\beta = N$ so that 
\begin{eqnarray}
\beta E(\{x_i\})=\frac{N^2}{2} \sum_{i=1}^N x_i^2 - \alpha N\, \sum_{i \neq j} |x_i-x_j| \;. \label{E-1dc_2}
\end{eqnarray}
This is the so called ``jellium'' model, whose bulk properties in the thermodynamic limit ($N \to \infty$) have been
studied extensively before~\cite{ Lenard61, Prager62, Baxter63, Dean10,TT2015}. In this paper, our focus is on the edge behaviour, in particular the distribution of the position of the rightmost particle $x_{\max}$, as well as the gap $g$ between the positions of the two rightmost charges. In addition, we also compute the distribution of the index $N_+$ in the large $N$ limit. Let us summarise our main results:   
\begin{itemize}
\item[a)]{Distribution of $x_{\max}$: It is well known that, in the large $N$ limit, the average density of charges (normalised to unity) is uniform $\rho_N(x)|_{N\to \infty} = \rho_\infty(x) =  \frac{1}{4\alpha}$, for $-2\alpha \leq x \leq 2 \alpha$. Thus the average $\langle x_{\max}\rangle \to 2 \alpha$. For large but finite $N$, $x_{\max}$ fluctuates around this mean value with typical fluctuations scaling as $O(1/N)$. Indeed we compute the full cumulative distribution 
\begin{equation}
Q(w,N)=\text{Prob}.\,[x_{\max}\le w, N], \label{Q-1dc}
\end{equation}
and show that it exhibits three different regimes (see Fig. \ref{fig1})
\begin{align}
\small
Q(w,N) \approx
\begin{cases}
&\ee^{- N^3\, \Phi_{-}(w) + O(N^2)},~~~~~~~~0< 2\al-w \sim O(1) \\
& \\
&F_\al(N(w-2\alpha) + 2 \alpha),~~~|w-2\al| \sim O(1/N)\\
&\\
&1- \ee^{- N^2\, \Phi_+(w)+O(N)},~~~~~ 0<w-2\al \sim O(1) \;.
\end{cases}
\label{result-1}
\end{align}
The second line denotes the regime for typical fluctuations $|w - 2 \alpha| \sim O(1/N)$ where the scaling function $F_\alpha(x)$ satisfies a nonlocal eigenvalue equation 
\begin{equation}
\frac{dF_\alpha(x)}{dx}= A(\al)\, \ee^{-x^2/2}\, F_\alpha(x+4\al) \;,
\label{limit.00}
\end{equation}
with $A(\al)$ as the unique eigenvalue that can be determined (see later). The tails of the distribution $F_\al(x)$ are given by 
\begin{eqnarray}
F_\al'(x) \approx
\begin{cases}
&\exp\left[-|x|^3/{24\al}+ O(x^2)\right ] \; \textrm{as}\,\, x \to -\infty \\
& \vspace*{-0.25cm} \\
& \exp\left[-x^2/2 + O(x)\right] \hspace*{0.8cm}
\,\textrm{as}\,\, x \to \infty \;.
\end{cases}
\label{fx_asymp-0}
\end{eqnarray}  
This is the analogue of the TW distribution found in the log-gas case (see Eq. (\ref{regimes_gaussian_CDF})). Clearly, this limiting distribution is different from the TW distribution.  

The first and third lines in Eq. (\ref{result-1}) describe the atypical large fluctuations of $x_{\max}$, respectively to the left and the right of the central typical regime. We compute explicitly the rate functions $\Phi_-(x)$ and $\Phi_+(x)$:  
\begin{eqnarray}
\label{sm-left_rf-0}
\Phi_{-}(w) = 
\begin{cases}
& \frac{(2\al-w)^3}{24\al}\;,\;\quad -2\al\leq w \leq 2\al \\
&Ê\vspace*{-0.3cm} \\
& \frac{w^2}{2} + \frac{2}{3} \al^2 \;, \;\quad\; w \leq -2\al \;.
\end{cases}
\end{eqnarray}
\begin{eqnarray}
\label{sm-phi+0}
\Phi_+(w) = \frac{(w-2\al)^2}{2} \;, \quad\;\;\;\quad w > 2\al \;.
\end{eqnarray}
The three regimes in Eq. (\ref{result-1}) are shown schematically in Fig. \ref{fig1}.
It is easy to check that the central part described by $F_\al(x)$ in Eq. (\ref{result-1}) matches smoothly 
 with the two large deviation regimes flanking this central part.}
\item[b)]{Distribution of the gap: we compute the distribution $P_G(g,N)$ of the gap $g$ between the positions of the two rightmost particles. We show that, for large $N$, it has the scaling form
\begin{eqnarray}\label{eq:scaling_g}
P_G(g,N) \approx N~h_\alpha(gN) \;,
\end{eqnarray}
where the scaling function $h_\alpha(z)$ is given by
\begin{eqnarray}\label{eq:halpha}
h_\alpha(z) &=& \Theta(z)\,A(\alpha) \int_{-\infty}^\infty dy~(y+z-4\alpha)~e^{-(y+z-4\alpha)^2/2}~F_\alpha(y) \;.
\end{eqnarray}
In Eq. (\ref{eq:halpha}), $F_\alpha(x)$ is again the unique solution of Eq. (\ref{limit.00}) with eigenvalue $A(\alpha)$ and $\Theta(z)$ is the Heaviside step function.  
}
\item[c)]{Distribution of the index: we have also computed analytically, for large $N$, the distribution $P_I(N_+,N)$ of the index $N_+$, i.e. the number of charges $N_+= \sum_{i=1}^N \theta(x_i)$ on the positive semi-axis. From the symmetry of the energy $E(\{x_i\})$ about the origin in \eqref{E-1dc}, it is evident that $\langle N_+ \rangle =N/2$ and furthermore the full distribution $P_I(N_+,N)$ is symmetric around $N_+ = N/2$. Indeed, for large $N$, 
we show that it approaches a scaling form
\begin{equation}
P_I(N_+, N) \approx 4\alpha\, f_\alpha\left( 4 \alpha \left(N_+-{N}/{2}\right)\right)\, ,
\label{limiting.1_intro}
\end{equation}
where the scaling function $f_\alpha(z)$ is given by
\begin{equation}
f_\alpha(z)=  \frac{F_\alpha(z+2\alpha)\, F_\alpha(-z+2\alpha)}{\int_{-\infty}^{\infty} dz\, 
F_\alpha(z+2\alpha)\, F_\alpha(-z+2\alpha)} \;.
\label{limiting.2_intro}
\end{equation}
where $F_\alpha(x)$ is again the unique solution of (\ref{limit.00}). Note that the typical fluctuations of $N_+$ around its mean 
are here of order $O(1)$ [see Eq.~(\ref{limiting.1_intro})], while they are of order $O(\sqrt{\ln N})$ in the log-gas [see Eq.~(\ref{gauss_index})]. Moreover, we show that the scaling function 
$f_\alpha(z)$ has non-Gaussian tails  
\begin{equation}
f_\alpha(z) \sim \exp\left[- \frac{1}{24\alpha}\,|z|^3\right]\quad {\rm as}\quad |z|\to \infty \, .
\label{asymp.1_intro}
\end{equation}
This limiting distribution is thus clearly non-Gaussian, unlike in the log-gas case where it is known to be Gaussian [see Eq. (\ref{gauss_index})]. The function $f_\alpha(z)$ describes only the typical fluctuations of $N_+$ of order $O(1)$ around its mean. 
The atypical fluctuations of order $O(N)$ on both sides of the mean are described 
 by symmetric large deviation tails:
 \begin{equation}
P_I(N_+=cN,N)\approx \exp \left(-N^3 \, \Psi(c) \right) \;\;, \;\; \Psi(c) = \frac{8 \alpha^2}{3}~|c-1/2|^3 \;,~~~0\le c \le 1 \;. \label{LDF}
\end{equation}  
The rate function $\Psi(c)$ is simple here and is rather different from the corresponding one in the log-gas case~\cite{Majumdar09b,Majumdar11}). 
}
\end{itemize}

\section{Some basics on the $1d$ jellium model}\label{P_x_max}

With the choice $\beta = N$, the partition function of the model is given by
\begin{eqnarray} 
\hspace*{-2.2cm}Z_N = \int_{-\infty}^\infty \e^{- \beta E(\left\{ x_i\right\})} dx_1 \cdots dx_N \;, \;{\rm with}\;\quad  \beta E(\left\{ x_i\right\}) = \frac{N^2}{2} \sum_{i=1}^N x_i^2 - \alpha N\, \sum_{i \neq j} |x_i-x_j| \;. \nonumber \\
\label{partition_func}
\end{eqnarray}
In the large $N$ limit (equivalently the zero temperature limit), the partition function is dominated by the ground state (minimum energy configuration). To find
this minimum energy configuration it is convenient to first rewrite the partition function $Z_N$ in (\ref{partition_func}) using the fact that $E(\{x_i \})$ is 
symmetric under permutations of the $x_i$'s. Hence 
\begin{eqnarray} 
Z_N = N ! \int_{-\infty}^\infty \cdots  \int_{-\infty}^\infty  dx_1 \cdots dx_N \; \e^{- \beta E(\left\{ x_i\right\})} \prod_{j=2}^N \Theta(x_j - x_{j-1})  \;.
\label{partition_func2}
\end{eqnarray}
For such an ordered configuration $x_1<x_2<\cdots<x_N$, we can eliminate the absolute values and rewrite the energy function as
\begin{eqnarray}
\beta E(\{x_i\})&=&\frac{N^2}{2} \sum_{i=1}^N x_i^2 - 2\alpha N \sum_{i > j} (x_i-x_j), \nonumber \\ 
&=&\frac{N^2}{2} \sum_{i=1}^N x_i^2 - 2\alpha N \sum_{i=1}^N (2i-N-1)x_i, \nonumber \\
&=&\frac{N^2}{2} \sum_{i=1}^N \left( x_i^2 - \frac{4\alpha}{N}(2i-N-1)x_i \right), \nonumber \\
&=& \frac{N^2}{2} \sum_{i=1}^N \left( x_i - \frac{2 \alpha}{ N}( 2i-N-1)\right)^2 + C_N(\alpha), 
\label{E-ordered}
\end{eqnarray} 
where $C_N(\alpha)= 2 \alpha^2 \sum_{i=1}^N \left( 2i-N-1\right)^2$ is just a constant. Clearly the minimum energy configuration corresponds to
\begin{equation}
x_i = x_i^*=\frac{2 \alpha}{N}(2i-N-1),~~~\text{for}~~i=1,2,...,N\label{x_j-min}
\end{equation}
This implies that, in the minimum energy configuration the charges are placed at regular intervals of length $\frac{4 \alpha }{N}$. The rightmost particle is at $x_N^* = 2 \alpha(1-1/N)$ and the leftmost particle is at the symmetrically opposite place $x_1^* = -2 \alpha(1-1/N)$. Hence it is clear that the 
charge density is supported over a finite support and in the large $N$ limit it is given by
\begin{equation}
 \rho_\infty(x) =
\begin{cases}
& \frac{1}{4\alpha},~~\text{for}~~-2\alpha \leq x \leq 2 \alpha \\ 
& 0~~~~~~~~~~~~~~~~~~~~~\text{otherwise}
\end{cases}. \label{rho-1dc}
\end{equation}
Note that $\rho_\infty(x)$ is different from the Wigner semi-circle \eqref{wig-sc} obtained in the log-gas.

For this jellium model, different thermodynamic properties have been studied extensively \cite{Choquard1981,Lenard61,Prager62,Baxter63,Dean10,TT2015}. In particular, Baxter \cite{Baxter63}
analysed the partition function $Z_{N,L}$ of the jellium model confined in a finite box $[-L,L]$, i.e. the following multiple integral 
\begin{eqnarray}\label{eq:ZNL}
Z_{N,L} = N ! \int_{-L}^L \cdots  \int_{-L}^L  dx_1 \cdots dx_N \;  \e^{- \beta E(\left\{ x_i\right\})} \prod_{j=2}^N \Theta(x_j - x_{j-1}) \;.
\end{eqnarray} 
In computing this integral (\ref{eq:ZNL}), Baxter introduced \cite{Baxter63}, as an intermediate step, an auxiliary function $F_\alpha(x)$ that satisfies 
a non-local eigenvalue equation defined in Eq.~(\ref{limit.00}). In this paper, we are interested in the distribution of three basic observables in an infinite
system: (i) the position of the rightmost particle $x_{\max}$, (ii) the gap $g = x_{N} - x_{N-1}$ between the positions of the two rightmost particles and (iii) the 
index $N_+$, i.e., the number of particles on the positive semi-axis. The distributions of these observables have not been studied in the classical
literature on the jellium model, to the best of our knowledge. Remarkably, we find that the same auxiliary $F_\alpha(x)$ function that Baxter introduced 
for the analysis of the partition function of the system in a finite box, also plays a key role in determining the distributions of these three observables
on the infinite line.

\section{Distribution of $x_{\max}$}\label{sec:max}

In this section, we focus on the position $x_{\max}$ of the rightmost particle on the infinite line. Clearly 
$x_{\max}$ is a random variable which fluctuates from one realisation to another. From the analysis of the average density in Eqs. (\ref{x_j-min}) and (\ref{rho-1dc}),
it is clear that, in the limit $N \to \infty$, the mean position of the rightmost particle is  
\begin{eqnarray}\label{av_xmax}
\langle x_{\max}\rangle = x_N^* \approx 2 \alpha \;. 
\end{eqnarray}
To derive the distribution of $x_{\max}$, it is convenient to consider the cumulative distribution \[Q(w,N) = \text{Prob}.[x_{\max} \leq w] = \text{Prob.}(x_1\leq w, \cdots,x_N\leq w).\] Using the Boltzmann distribution $\mathcal{P}(x_1,x_2,...,x_N)$ in Eq. (\ref{jpdf-1dc}), one can express $Q(w,N)$ as the ratio of two partition functions  
\begin{eqnarray}
Q(w,N)&=& \frac{Z_N(w)}{Z_N(\infty)},~~\text{where,}\label{Q_N-1} \\
Z_N(w)&=& \int_{-\infty}^w dx_1\cdots \int_{-\infty}^w dx_N~\ee^{-\beta E(\{x_i\})}, \label{Z_N-w}
\end{eqnarray}
where $\beta\,E(\{x_i\})$ is given in \eqref{E-1dc} and $Z_N(\infty) \equiv Z_N$ given in Eq. (\ref{partition_func}). Again, it is convenient to work with
ordered configurations of the $x_i$'s, $-\infty < x_1\leq x_2 \leq...\leq x_N \leq w$ as before and one gets (using Eq. (\ref{E-ordered}))
\begin{eqnarray}
Z_N(w) \propto \int_{-\infty}^w d x_N \int_{-\infty}^{x_N} d x_{N-1}...\int_{-\infty}^{x_2} d x_{1}~\ee^{-  \frac{N^2}{2} \sum_{i=1}^N \left( x_i - \frac{2 \alpha}{ N}( 2i-N-1)\right)^2}
\label{Z_N-w-ord}
\end{eqnarray}
where we have replaced the product of theta functions by constraining the limits of the integrals. It is natural now to make a change of variables 
\begin{equation}
\ep_i=\left[Nx_i-2\al(2i-N-1)\right],~~i=1,2,\cdots,N \;. \label{transformation}
\end{equation} 
The ordering condition $x_{i-1} < x_i$ translates to the following constraint on the $\epsilon_i$'s
\begin{eqnarray}\label{cond_epsilon}
\ep_{i-1}<\ep_i+4 \alpha ,~~\text{for}~i=2,3,...,N \;.
\end{eqnarray}
The last constraint $x_N < w$ in Eq. (\ref{Z_N-w-ord}) translates to 
\begin{eqnarray}\label{cond_epsilon_N}
\epsilon_N < N\,(w-2\al)  + 2 \alpha  \;.
\end{eqnarray}
Consequently, $Z_N(w)$ reads
\begin{equation}
Z_N(w) \propto D_\alpha\left(N(w-2\al)+2\al,N\right), \label{Z_N_D_a}
\end{equation}
where  the function $D_\al(x,N)$ on the right hand side (rhs) is given by the $N$-fold integral 
\begin{align}
D_\al(x,N) &=~\int_{-\infty}^{x} d \ep_N\int_{-\infty}^{\ep_N+4 \al}d \ep_{N-1}\dots \int_{-\infty}^{\ep_2+4\al} d\ep_1~\ee^{-\frac{1}{2}\sum_{i=1}^N \ep_i^2} \;.  \label{D_N} 
\end{align}
We remark that in the original jellium model in Eq. (\ref{partition_func}), the interaction between the $x_i$'s
is long-ranged (as every charge is coupled to every other charge). Remarkably however, after the ordering 
of the positions and the change of variables in Eq. (\ref{transformation}), the interactions between the
new variables $\epsilon_i$'s in Eq. (\ref{D_N}) become short-ranged, i.e., $\epsilon_i$ interacts only with its two nearest
neighbours $\epsilon_{i-1}$ and $\epsilon_{i+1}$. Therefore, the function $D_\al(x,N)$ in Eq. (\ref{D_N}) can be interpreted
as a restricted partition function of this constrained short-ranged interacting gas.

To make further progress, we substitute $Z_N(w)$ from Eqs. (\ref{Z_N_D_a}) and (\ref{D_N}) into Eq.~(\ref{Q_N-1}) and obtain
\begin{eqnarray}
Q(w,N) = \frac{D_\al(x,N)}{D_\al(\infty,N)} \equiv F_\alpha(x,N) \;. \label{def_FN}
\end{eqnarray}
Taking derivative with respect to $x$ in  (\ref{def_FN}), and using  (\ref{D_N}), we obtain
\begin{equation}\label{dF_N.1}
\frac{d\,F_\al(x,N)}{d\,x} = \frac{D_{\al}(\infty,N-1)}{D_{\al}(\infty,N)}\,\ee^{-\frac{x^2}{2}} F_{\al}(x+4\al,N-1) \;.
\end{equation}
These equations (\ref{def_FN}) and (\ref{dF_N.1}) are actually exact for all $x$ and $N$. To make progress, we will consider the
$N \to \infty$ limit. In this limit, the typical fluctuations of $x_{\max}$ around its mean value $2 \alpha$ turn out to be of order $O(1/N)$, while atypical large
fluctuations can be of order $O(1)$. Below, we analyse the probability distribution of typical and atypical fluctuations separately. 

\subsection{Typical fluctuations of $x_{\max}$}

To analyse the typical fluctuations, we need to keep the argument $x = (w-2\al) N + 2 \al$ of $D_\alpha(x,N)$ fixed in Eq. (\ref{Z_N_D_a}), while we take the $N \to \infty$ limit. In addition, we need to estimate the ratio $\frac{D_{\al}(\infty,N-1)}{D_{\al}(\infty,N)}$ in Eq. (\ref{dF_N.1})
in the large $N$ limit. As discussed earlier, since $D_\al(\infty,N)$ is the partition function of a short-ranged gas, we expect that
its free energy $-\ln D_\al(x,N)$ is extensive in $N$. Hence it follows that $D_\al(\infty,N) \sim [A(\alpha)]^{-N}$ for large $N$, where $\ln A(\alpha)$ is the free energy per particle of the short-ranged interacting gas. Hence the ratio 
\begin{equation}
\frac{D_{\al}(\infty,N-1)}{D_{\al}(\infty,N)} \to A(\alpha),~~\text{as}~~N \to \infty \;.
\label{D-inf}
\end{equation} 
We substitute this result (\ref{D-inf}) on the rhs of Eq. (\ref{dF_N.1}). Anticipating further that 
the function $F_\al(x,N)$ converges to a limiting form $F_\al(x)$ for large $N$, {i.e.}
\begin{align} 
F_\alpha(x,N \to \infty) = F_\alpha(x)\;, \label{limit.0} 
\end{align}
we find that $F_\al(x)$ satisfies a nonlocal equation 
\begin{equation}
\frac{dF_\alpha(x)}{dx}= A(\al)\, \ee^{-x^2/2}\, F_\alpha(x+4\al) \;.
\label{limit.1}
\end{equation}
The prefactor $A(\al)$ on the rhs is still unknown. The function $F_\alpha(x)$ is a cumulative probability distribution and hence
satisfies the positivity condition $0\le F_\alpha(x)\le 1$ for $-\infty < x < \infty$, along with the boundary conditions $F_\alpha(-\infty)=0$ and $F_\al(\infty)=1$. 
It turns out that the solution of Eq. (\ref{limit.1}) satisfies these conditions only for a specific value $A(\al)$ -- in this sense Eq. (\ref{limit.1}) can be interpreted as a non-local eigenvalue equation. 

As we have remarked earlier, the same non-local eigenvalue equation (\ref{limit.1}) also appeared in Baxter's analysis
of $Z_{N,L}$ in Eq. (\ref{eq:ZNL}). Indeed, we remark that we can give a probabilistic interpretation to the integral 
$Z_{N,L}$ in Eq. (\ref{eq:ZNL}). Up to a prefactor, this is just the probability that all the particles in an infinite system
are contained in $[-L,+L]$, which in turn, is the probability that the maximum of $|x_i|$'s is less than $L$, i.e.
\begin{eqnarray}\label{Z_NL_prob}
Z_{N,L} \propto {\rm Prob.}\, \left[\max\{|x_1|, |x_2|, \cdots, |x_N|\} < L\right] \;.
\end{eqnarray}  

\begin{figure}[t]
  \centering
    \includegraphics[width=\linewidth]{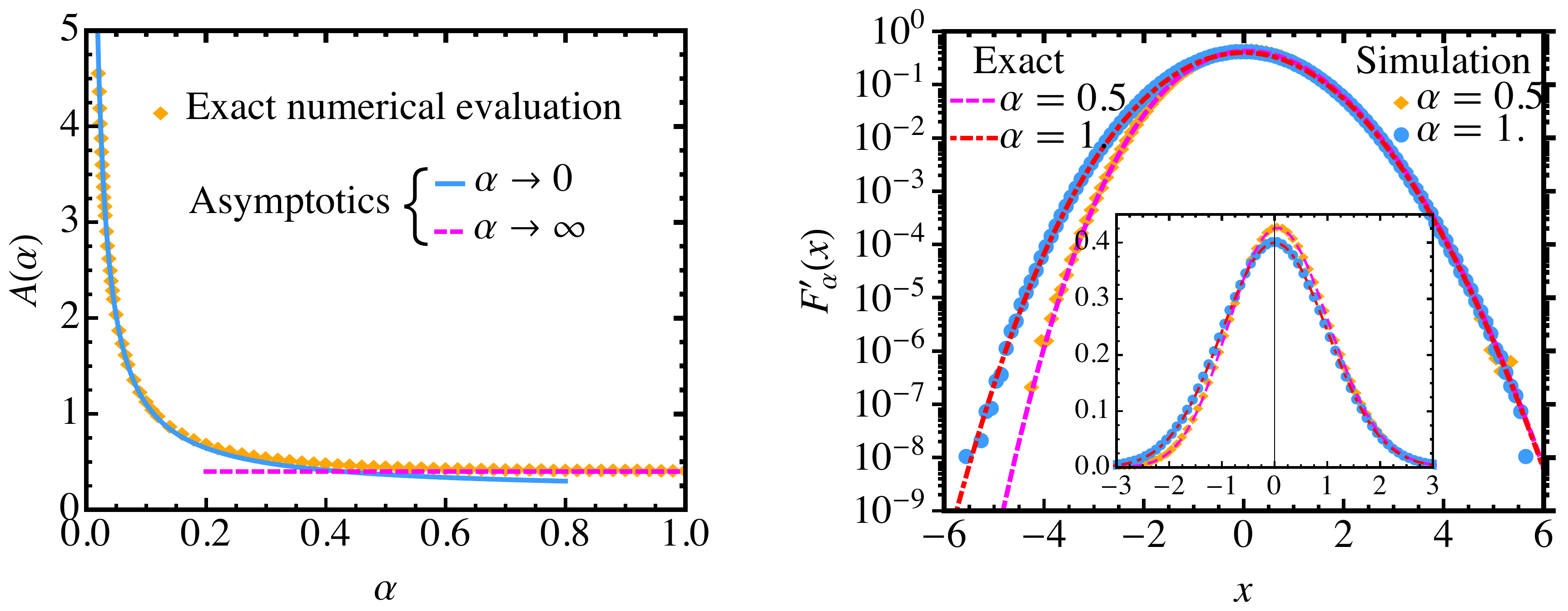}
  \caption{(Left): Plot of $A(\al)$ and numerical verification of its $\al \to 0$ and $\al \to \infty$ asymptotic. (Right): Comparison of the theoretical $F_\al'(x)$ obtained by solving numerically \eqref{limit.1} by a shooting method and $F_\al'(x)$ obtained from direct Monte-Carlo simulation of the ``jellium'' model (with $N=50$) for two different values of the coupling parameter $\alpha = 1$ and $\alpha = 0.5$. Inset shows the distribution in the normal scale.} 
\label{A-al-fig}
\end{figure} 

While computing $A(\alpha)$ analytically for all $\alpha>0$ seems hard, it is possible to determine its small and large $\al$ behaviours. For large $\al$, one can simply replace the upper limit of the integrals in $D_\al(\infty,N)$ in Eq. (\ref{D_N}) by $\alpha \to \infty$. This gives $D_\al(\infty, N) = (\sqrt{2\pi})^N$. Consequently, the ratio $A(\al \to \infty) \to 1/\sqrt{2 \pi}$. In contrast, the $\alpha \to 0$ limit is less trivial. However, this was already determined by Baxter \cite{Baxter63}. Translating his asymptotic results to our case (his notations are quite different from ours) we finally get
\begin{equation}
A(\al) \to 
\begin{cases}
&1/(4\,\e\,\al),~~\text{as}~~\al \to 0 \\
&1/\sqrt{2 \pi},~~~~~\text{as}~~\al \to \infty.
\end{cases}
\label{a-al-asymp}
\end{equation}
For other values of $\alpha$, $A(\alpha)$ can be computed numerically using a shooting method, as
discussed later. 

\noindent{\it Asymptotic behaviours of $F_\alpha(x)$.} The tails of $F_\alpha(x)$ to leading order can be determined for arbitrary $\alpha > 0$ as it does not require
the explicit knowledge of $A(\alpha)$. We first consider the $x \to \infty$ limit. In this limit we replace $F_\alpha(x + 4 \alpha)$ by 1 on the
rhs of \eqref{limit.1}. This gives, to leading order, the Gaussian tail $F_\alpha'(x) \sim \ee^{-x^2/2}$ for large $x$. To compute the left tail, $x \to -\infty$, 
we make the following ansatz  
\begin{equation}\label{Falpha-inf}
F_\alpha(x) \approx \ee^{-a_0\,|x|^\delta},~~\text{as}~x \to -\infty,
\end{equation} 
where $a_0$ and $\delta$ are to be determined. 
We substitute this ansatz on both sides of \eqref{limit.1} and then equate the powers of $|x|$ in the exponential. The rhs yields $rhs \approx A(\al)\,\ee^{-x^2/2-a_0\,(|x|-4\al)^\delta}$.
For large $|x|$, $(|x|-4\al)^\delta \sim |x|^\delta(1 - 4\,\alpha\,\delta/|x|)$ to leading orders. Hence the rhs behaves as 
\[ \text{rhs} \approx A(\al) \ee^{-a_0\,|x|^\delta - x^2/2 + 4\,\al\,\delta\,|x|^{\delta-1}}. \]
The left hand side (lhs) of  (\ref{fx_asymp}) behaves as 
\[ \text{lhs}\approx \ee^{-a_0\, |x|^\delta} \]
to leading order. Comparing both sides, we see that the term $x^2/2$ and $|x|^{\delta-1}$ on the rhs must cancel each other. This implies that $\delta = 3$ and $a_0 = 1/(8\alpha \,\delta) = 1/(24\, \alpha)$. This provides the leading left tail $F_\alpha'(x)$ in \eqref{fx_asymp}.  Together, the leading order asymptotic tails are given by
\begin{eqnarray}
F_\al'(x) \approx
\begin{cases}
&\exp\left[-|x|^3/{24\al}+ O(x^2)\right ] \; \textrm{as}\,\, x \to -\infty \\
& \vspace*{-0.25cm} \\
& \exp\left[-x^2/2 + O(x)\right] \hspace*{0.8cm}
\,\textrm{as}\,\, x \to \infty \;.
\end{cases}
\label{fx_asymp}
\end{eqnarray}   
Note that the leading left tail of $F'_\al(x)$ is similar to the left tail of the TW distribution in \eqref{TW-asymptotic} (with $\beta = 1/\alpha$), while the right tail in Eq. (\ref{fx_asymp}) is different from the right tail in Eq. (\ref{TW-asymptotic}). 

For general $\al>0$, it is difficult to determine the eigenvalue $A(\alpha)$ as well as the full scaling function $F_\alpha(x)$ explicitly. However they can be obtained by solving \eqref{limit.1} numerically by tuning the value of $A(\al)$ using the standard shooting method \cite{shooting}. This gives $F_\al(x)$ and $A(\al)$ simultaneously. In Fig.~\ref{A-al-fig} (left panel), we plot $A(\al)$ {vs.} $\al$ and compare with its predicted asymptotics in \eqref{a-al-asymp}. In Fig.~\ref{A-al-fig} (right panel), we compare $F_\al'(x)$ evaluated numerically using this shooting method, with the one obtained from direct Monte-Carlo simulation of the jellium model. The agreement is excellent.

\subsection{Atypical large fluctuations of $x_{\max}$}\label{LDF-xmax}


In the previous section we have studied the typical fluctuations of $x_{\max}$ on a scale of order $O(1/N)$ 
around its mean $2\al$ in the large $N$ limit. We have shown that this centred and scaled limiting cumulative 
distribution is described by ${\rm Prob.}(x_{\max}<w) = Q(w,N) = \approx F_\alpha(N(w-2\al) + 2\al)$ where the scaling
function $F_\alpha(x)$ is given in Eq. (\ref{limit.1}), along with the tails given in Eq. (\ref{fx_asymp}). However this limiting 
distribution does not describe large fluctuations of $O(1)$ at far left or right of the mean. In the log-gas case, these large deviation
functions were computed exactly \cite{Majumdar14} as described in the introduction, that revealed an interesting third order phase transition between
a ``pushed'' and a ``pulled'' phase. It is then interesting to ask whether a similar phase transition also exists in the jellium model. 
This motivated us to study the probability of large deviations in the jellium model. Our exact computations show that a similar third order 
phase transition also exists in this case. Below, we discuss the left and right large deviation functions separately as they correspond 
to different physics.

\subsubsection{Left large deviation:}
\begin{figure}[t]
\centering\includegraphics[scale=0.35]{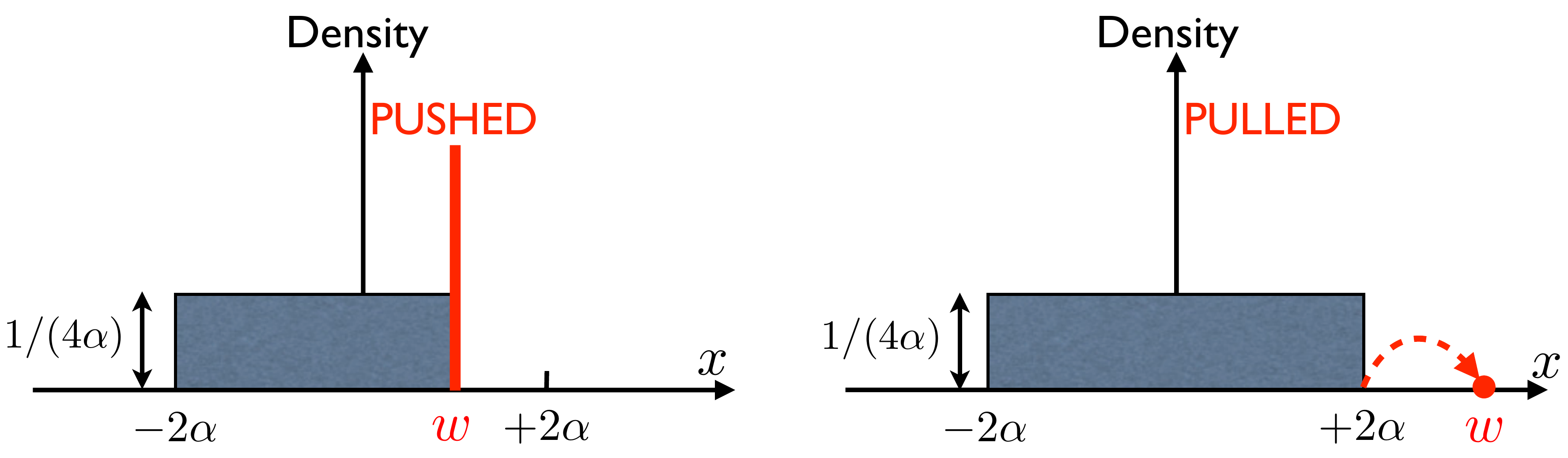}
\caption{{\bf Left:} the left large deviation function $\Phi_-(w)$ [see  (\ref{sm-left_rf})] is obtained by computing the free energy cost in {\it pushing} the wall to the left of the right edge, i.e. $w < 2\al$, of the flat equilibrium density. {\bf Right:} the right rate function $\Phi_+(w)$ [see  (\ref{sm-phi+})] is evaluated by computing the energy cost in {\it pulling} a single charge  at $0< w-2\al \sim O(1)$ from the flat equilibrium distribution of charges.}
\label{Fig_pushed_pulled}
\end{figure}

We start with Eqs. (\ref{Q_N-1}) and (\ref{Z_N-w}), with the energy $\beta E(\{ x_i\})$ given in Eq. (\ref{partition_func}).  We need to compute the leading behaviour of  the partition function $Z_N(w)$ for large $N$ with a wall at $w$ such that $0 < (2 \alpha - w) \sim 1$. This can be performed as follows:
One first introduces a macroscopic empirical charge density in $(-\infty,w]$ 
\begin{equation}\label{sm:rho_w}
{\rho}_w(x)= \frac{1}{N} \sum_{i=1}^N \delta(x-x_i) \;.
\end{equation}
Note that $\rho_w(x)$ is normalised to unity. In terms of $\rho_w(x)$, the energy function $\beta E(\{x_i \})$ in Eq. (\ref{partition_func}) can be expressed as 
\begin{equation}
\beta\,E[\{x_i\}] \equiv \mathcal{E}[\rho_w(x)]
\end{equation}
where
\begin{equation}\label{sm-E-Ncube}
\mathcal{E}[\rho_w(x)]=N^3\,\left(\frac{1}{2} \int_{-\infty}^w dx~x^2\rho_w(x) - \al\int_{-\infty}^w dx\int_{-\infty}^w dy~\rho_w(x)\rho_w(y)~|x-y|\right) \;.
\end{equation}
The $N$-fold integration in the partition function $Z_N(w)$ in  (\ref{Z_N-w}) is carried out in two steps. In the first step, we fix the macroscopic density $\rho_w(x)$ and then sum over all the microscopic configurations of $x_i$'s consistent with this density $\rho_w(x)$. In the second step, we sum over all possible macroscopic densities $\rho_w(x)$ that are positive and normalised to unity $\int_{-\infty}^w \rho_w(x) \,dx= 1$. The first step gives rise to an entropy term that scales, for large $N$, as $O(N)$ (see for instance \cite{Dean08}). But since the energy $\mathcal{E}[\rho_w(x)]$ in Eq. (\ref{sm-E-Ncube}) scales as $N^3$, we can neglect the entropy term at leading order for large $N$. This gives
\begin{equation}
Z_N(w)\approx \int \mathcal{D}\rho_w ~\text{exp}\left (- \beta \mathcal{E}[\rho_w(x)]\right)\,\delta \left(\int_{-\infty}^w dx~\rho_w(x)-1 \right) \;. \label{partition-2}
\end{equation}
where $\mathcal{D}\rho_w$ denotes the measure of a functional integral over all possible densities satisfying the normalisation constraint $\int_{-\infty}^w dx~\rho_w(x) = 1$. To proceed further we replace the delta function by its integral representation and get 
\begin{align}
Z_N(w)=&N^3 \int \frac{d \mu}{2 \pi i} \int \mathcal{D}\rho_w ~\text{exp}\left (- N^3 \,\mathcal{S}[\rho_w(x)] \right), ~~\text{with}\label{partition-3} \\
\mathcal{S}[\rho_w(x)]=& \left[ \frac{1}{2} \int_{-\infty}^w dx~x^2\rho_w(x) - \al  \int_{-\infty}^w dx\int_{-\infty}^w dy~\rho_w(x)\rho_w(y)~|x-y| \right. \nonumber \\ 
&~~~~~~~~~~~~~~~~~~~~~~~~~~~~~~\left. +~\mu \left(\int_{-\infty}^w dx~\rho_w(x)-1 \right) \right] \;. \label{action}
\end{align}
The integral in \eqref{partition-3} can be performed, for large $N$, by a saddle point approximation that gives
\begin{equation}\label{Z_saddle}
Z_N(w) \approx \text{exp}\left (- N^3 S[\rho^*_w(x)] \right),
\end{equation}
where $\rho^*_w(x)$ is the saddle point density that minimises the action $S[\rho_w(x)]$ in (\ref{action}). 
The equation for $\rho^*_w(x)$ is obtained from $\left(\frac{\delta S[\rho_w]}{\delta \rho_w(x)}\right)_{\rho_w=\rho^*_w}=0$ as 
\begin{equation}
\frac{1}{2}x^2 -2\alpha \int_{-\infty}^w dy~\rho^*_w(y)~|x-y| +\mu = 0 \;. \label{sadl-eq-1}
\end{equation}
This equation holds for $ -\infty \le x \le w$. But, the support of $\rho_w^*(x)$ can not extend over the full range $(-\infty,w]$ due to the following reasons: When $x \to -\infty$, the first term in  (\ref{sadl-eq-1}) grows as $x^2$, while the second term grows as $|x|$ -- hence they can not compensate each other. However the equation \eqref{sadl-eq-1} holds true. The only possible way this can happen is that $\rho^*_w(x)$ has a finite support, say over $[-B,x]$ where $B$ can be determined from the normalisation constraint 
\begin{equation}
\int_{-B}^{w} \rho^*_w(x) dx =1\;. \label{norml-1}
\end{equation}
Outside this region $\rho_w^*(x)$ is zero. 
For $x \in [-B,w]$, differentiating twice the saddle point equation (\ref{sadl-eq-1}) and using the identity $\frac{d^2}{dx^2}|x-y|=2\delta(x-y)$, it is easy to show that $\rho^*_w(x) = 1/(4 \alpha)$. Clearly, if $w > 2 \al$, the saddle point density is given by 
\begin{eqnarray}\label{rho_star_eq}
\rho_w^*(x) = \frac{1}{4 \al} \; \quad{\rm for} \; \quad-2\al \leq x \leq 2 \al \;, \quad \quad w >  2\al\;.
\end{eqnarray}
Thus for $w > 2\al$ the charge density does not change from its flat equilibrium density -- this is because the charges do not feel the presence of the wall. However, when $w<2\al$, the wall tries to push the charges to the left of $2\al$ (see the left panel of Fig. \ref{Fig_pushed_pulled}). We have seen from above that the bulk density does not change from its equilibrium value $\rho^*_w (x) = 1/(4\al)$ to the left of the wall at $w$. Normalisation to unity of the charge density then implies that the extra charge that the wall displaces must be accumulated at the wall, since the bulk is not affected. This leads, for $w<2\al$, to a new saddle point density of the form   
\begin{equation}
\rho^*_w(x)= \frac{1}{4\al}+ C~\delta(x-w),~\text{for}~-B\le x < w, \label{rho-guess}
\end{equation}
where $C$ represents the density of the charges displaced and absorbed at the wall. We have three unknowns: $B,~C$ and $\mu$ which are to be determined now.  From the normalisation condition $\int_{-B}^w \rho^*_w(x)\,dx = 1$ we get the relation between the two parameters $B$ and $C$ via
\begin{equation}
\frac{(w+B)}{4\al} + C=1 \;. \label{eq-3}
\end{equation}
We need two more equations. For that, we substitute the saddle point density $\rho^*_w(x)$ in (\ref{sadl-eq-1}) to get 
\begin{eqnarray}
\frac{1}{2}x^2 -\frac{1}{2} \int_{-B}^w dy~|x-y|-2 \al~C~ |x-w| +\mu = 0 \;. \label{eq-17-supp}
\end{eqnarray}
Now performing the integral over $y$ explicitly, we find
\begin{equation}
\left( 2\al~C + \frac{w-B}{2}\right)x ~+~\left(\mu - 2 \al~C~w-\frac{B^2+w^2}{4}\right)=0, ~\text{for}~-B\leq x <w \;. \label{eq-18-supp}
\end{equation}
Since  (\ref{eq-18-supp}) is valid for arbitrary $x \in [-B,w]$, the coefficients of different powers of $x$ are individually zero. As a result we get two additional equations 
\begin{eqnarray}
&&2\al\,C + \frac{w-B}{2}=0, \label{eq-1}\\ 
&&\mu = 2 \al~C~w+\frac{B^2+w^2}{4}. \label{eq-2}
\end{eqnarray}
We therefore have three equations (\ref{eq-3}), (\ref{eq-1}) and (\ref{eq-2}) for three unknowns $\mu, B$ and $C$, solving which we get
\begin{eqnarray}
B&=&2\al, \label{B} \\
C&=&\frac{1}{2} - \frac{w}{4\al}, \label{A} \\
\mu &=& \al^2 + \al w - \frac{w^2}{4} \;. \label{mu}
\end{eqnarray}
Since $B,~w$ and $\alpha$ are all positive by definition, then normalisation condition in \eqref{eq-3} implies $C \leq 1$. 
This indicates that the above analysis is valid only for $w > -2\alpha$. When $w \to -2\al$, $C \to 1$: this means that all the charges are absorbed at the wall and there is no bulk charge left. Thus for $w < -2\al$, we have effectively a single charge located at $w$ subjected to a harmonic potential. Therefore, for the saddle point density $\rho_w^*(x)$ we have the following expressions, valid for all $w$  
\begin{equation}
\label{rho_summary}
\rho^*_w(x) = 
\begin{cases}
&\frac{1}{4\al}\;,  \quad \hspace*{3.2cm}-2\al \leq x \leq 2\al \;, \quad\quad {\rm for} \quad\qquad\;\;\;\hspace*{0.32cm} w > 2\al \\
& \\
& \frac{1}{4\al} + \left(\frac{1}{2} - \frac{w}{4\al} \right)~\delta(x-w)\;, \quad -2\al\leq x \leq w \;, \quad {\rm for} \quad -2\al \leq w \leq 2\al \\
& \\
&\delta(x-w) \quad \quad \hspace*{5.7cm} {\rm for} \hspace*{1.7cm} w < -2\al \;.
\end{cases}
\end{equation}
This saddle point density $\rho^*_w(x)$ has a nice interpretation. When the wall position $w > 2\al$, it is given by the unperturbed density given in the first line of Eq. (\ref{rho_summary}) -- the charges do not feel the presence of the wall. When the wall position $-2\al<w<2\al$, the wall displaces
the charges over the region $[w,2 \al]$ and absorbs them on the wall as shown by the delta function term in the second line of Eq. (\ref{rho_summary}) (see also the left panel of Fig. \ref{Fig_pushed_pulled}). Finally, when $w<-2\al$, all the bulk charges are absorbed on the wall and the density is a simple delta function, given by the third line of Eq. (\ref{rho_summary}).

Our next task is to insert this saddle point density in the action $\mathcal{S}$ in \eqref{action} and get the partition function $Z_N(w)$ in \eqref{Z_saddle} to leading order. 
Let us first consider $w > 2 \al$. In this case $\rho_w^*(x) = 1/(4 \al)$ for $x \in [-2\al,+2\al]$. Substituting this density in  (\ref{action}) we get the saddle point action
\begin{eqnarray}
S[\rho_w^*(x)] = - \frac{2}{3}\alpha^2\;, \quad \quad {\rm for} \quad w > 2\al \;. \label{S_largew}
\end{eqnarray}
Therefore from  (\ref{Z_saddle}) the partition function $Z_N(w)$ for large $N$ and for $w > 2\al$ behaves as 
\begin{eqnarray}
Z_N(w) \approx \ee^{\frac{2}{3}\alpha^2\,N^3 } \;, \quad \quad {\rm for} \quad w > 2\al \;. \label{Z_largew}
\end{eqnarray}
In particular, taking $w \to \infty$ limit, we obtain the denominator in  (\ref{Q_N-1}) as
\begin{eqnarray}\label{denominator}
Z_N(\infty) \approx \ee^{\frac{2}{3}\alpha^2\,N^3} \;.
\end{eqnarray}
Hence, finally, for $w > 2\al$, to leading order for large $N$, we get
\begin{eqnarray}\label{Q_largew}
Q(w,N) = \frac{Z_N(w)}{Z_N(\infty)} \approx 1  \;, \quad \quad {\rm for} \quad w > 2\al \;.
\end{eqnarray}
To calculate the corrections to this leading order result, we need to consider the right large deviations function, that will be 
computed in the next section.  
Let us now consider the region where $-2\al \leq w \leq 2\al$. Substituting the saddle point density $\rho_w^*(x)$ from the second line of ~(\ref{rho_summary}) in  (\ref{action}) we get
\begin{equation}\label{S_inter}
S[\rho_w^*(x)] = - \frac{8\al^3+12\al^2w-6\al w^2+w^3}{24 \al} \;, \quad \quad {\rm for} \quad -2\al \leq w \leq 2\al \;.
\end{equation}
Substituting this result in  (\ref{Z_saddle}) and using the expression for the denominator in  (\ref{denominator}) 
we get
\begin{eqnarray}
\label{Q_mediumw}
Q(w,N) &=& \frac{Z_N(w)}{Z_N(\infty)} \approx \ee^{- N^3\Phi_-(w)}  \;,
\end{eqnarray}
where the large deviation function $\Phi_-(w)$ actually has a very simple expression
\begin{eqnarray}\label{Q_mediumw2}
\Phi_-(w) &=& \frac{(2\al - w)^3}{24\,\al}\;,~~\text{for} ~~  -2\al \leq w \leq 2\al \;.
\end{eqnarray}

\begin{figure}
\centering \includegraphics[width = 0.7\linewidth]{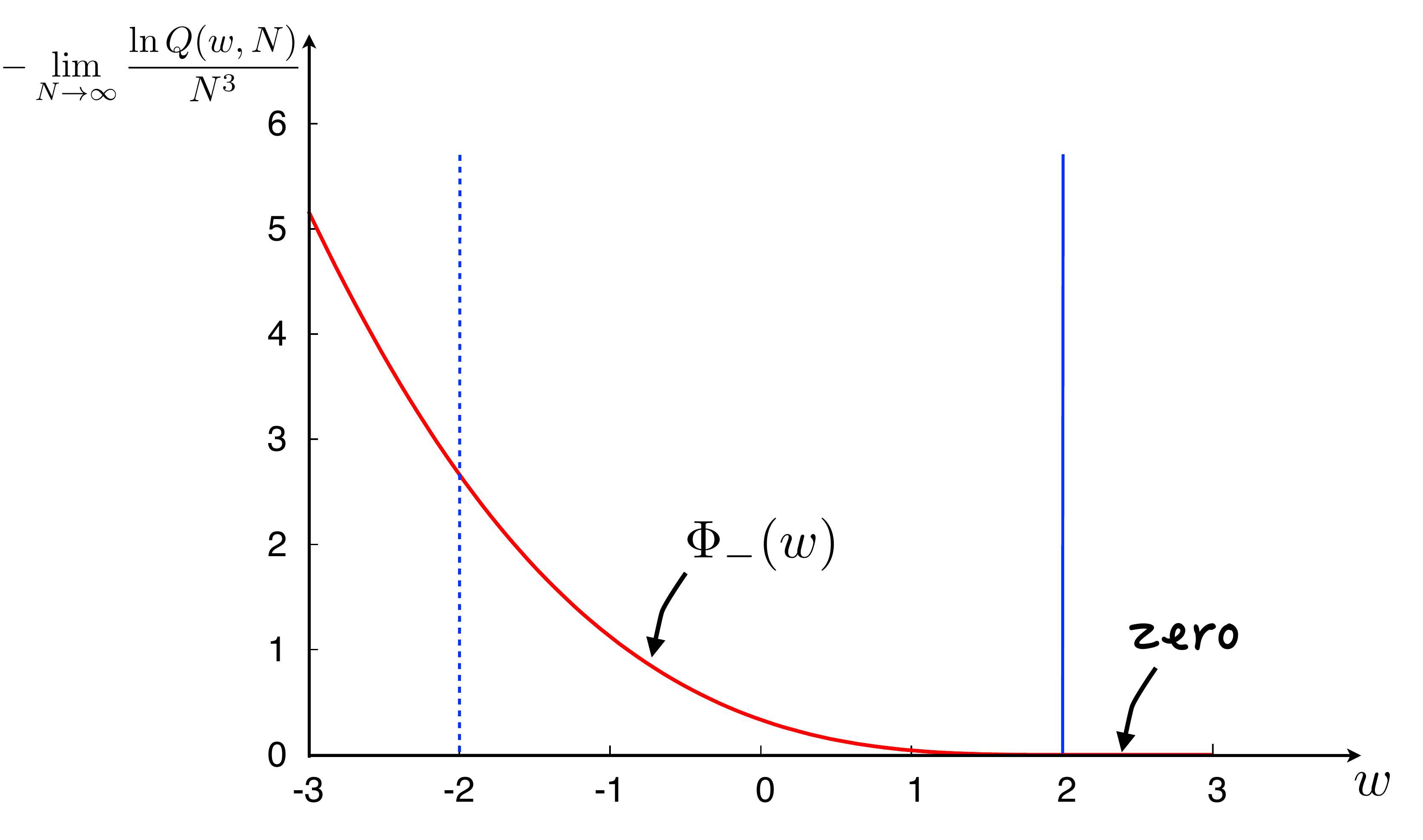}
\caption{Plot of the free energy $-\lim_{N \to \infty} {\ln Q(w,N)}/{N^3}$ as a function of $w$ for $\al=1$. For $w>2\al = 2$, the limiting value is just zero, $-\lim_{N \to \infty} {\ln Q(w,N)}/{\beta_0 N^2} = 0$, while it is non-zero for $w< 2 \al$, $-\lim_{N \to \infty} {\ln Q(w,N)}/{N^3} = \Phi_-(w)$ [see  (\ref{log_Qn})]. This transition at $w = 2\al = 2$ is indicated by the vertical solid blue line, where the third derivative of $\Phi_-(w)$ is discontinuous. This is the third order phase transition from the ``pulled'' ($w>2$) to the ``pushed'' ($w<2$) phase. The dotted vertical blue line at $w = - 2\al=-2$ (for $\al=1$) indicates another third order ''condensation'' transition when the pushed gas fully accumulates at the wall position.}\label{plot_phiminus}
\end{figure}

Finally, we consider the region where $w\leq-2\al$. In this case, 
substituting the saddle point density $\rho_w^*(x)$ from the third line of ~(\ref{rho_summary}) in  (\ref{action}) we get
\begin{eqnarray}\label{S_small}
S[\rho_w^*(x)] = \frac{w^2}{2}\;, \quad \quad {\rm for} \quad w < -2\al \;.
\end{eqnarray}
Substituting this result in  (\ref{Z_saddle}) and using the expression for the denominator in  (\ref{denominator}) 
we get
\begin{eqnarray}
\label{Q_smallw}
Q(w,N) &=& \frac{Z_N(w)}{Z_N(\infty)} \approx \ee^{-N^3\Phi_-(w)}  \;, \quad {\rm for}~\beta_0 \sim N \nonumber \\ 
\quad {\rm where,} && \Phi_-(w) = \frac{w^2}{2} + \frac{2}{3}\alpha^2\;,~~{\rm for}~~w\leq -2\al\;.
\end{eqnarray}
In summary 
\begin{eqnarray}\label{sm-left_rf}
\Phi_{-}(w) = 
\begin{cases}
& \frac{(2\al-w)^3}{24\al}\;,\;\quad -2\al\leq w \leq 2\al \\
&Ê\vspace*{-0.3cm} \\
& \frac{w^2}{2} + \frac{2}{3} \al^2 \;, \;\quad\; w \leq -2\al \;.
\end{cases}
\end{eqnarray}
In Fig. \ref{plot_phiminus}, we show a plot of $\Phi_-(w)$ as a function of $w$.

\vspace*{0.5cm}

\noindent{\it Third order phase transition at $w = 2\al$:} The cumulative distribution $Q(w,N)$ in  (\ref{Q_N-1}) is the ratio of two partition functions. Hence $-\ln Q(w,N) = -\ln Z_N(w)+\ln Z_N(\infty)$ can be interpreted as a free energy difference. Indeed, from ~(\ref{Q_largew}), we see that to leading order for large $N$, $-\ln Q(w,N) \approx 0$ for $w>2\al$. In contrast, for $w<2\al$, using  ~(\ref{Q_mediumw}) and (\ref{Q_smallw}), we see that $-\ln Q(w,N) \approx N^3 \Phi_-(w)$ where $\Phi_-(w)$ is given in  (\ref{sm-left_rf}). Hence, we get (see Fig. \ref{plot_phiminus})
\begin{eqnarray}\label{log_Qn}
- \lim_{N \to \infty} \frac{\ln Q(w,N)}{N^3} =
\begin{cases}
&0 \;, \;\hspace*{0.85cm} {\rm for}\; w > 2\al \\
&\Phi_-(w) \;, \; {\rm for}\; w < 2\al \;.
\end{cases}
\end{eqnarray}   
Thus $\Phi_-(w)$ is just the free energy cost in pushing the wall $w$ to the left of the right edge $2 \al$ (see Fig. \ref{Fig_pushed_pulled}). From the expression of $\Phi_-(w)$ in the first line of  (\ref{sm-left_rf}), it follows that $\Phi_-(w)$ vanishes as the third power $\Phi_-(w) \propto (2\al-w)^3$ as $w \to 2\al$ from the left. Thus the third derivative of the free energy is discontinuous at the critical point $2 \al$, making this a third order phase transition. Indeed, the pressure on the wall $P = -N^3 \Phi'_-(w)$ (derivative of the free energy with respect to the wall position) is zero for $w > 2\al$ (the charges do not touch the wall) and is non zero for $-2\al<w<2\al$. The mechanism of this third order transition is thus similar to the log-gas case \cite{Majumdar14}. 
However, in contrast to the log gas case, there is an additional third-order phase transition in the jellium model when $w \to -2\al$ [see Eq. (\ref{sm-left_rf})]. Indeed the third derivative of $\Phi_-(w)$ in Eq. (\ref{sm-left_rf}) is also discontinuous at $w = - 2 \al$. This transition is not of the ``pushed-pulled" type like the one at $w =2\al$, but rather a condensation-type transition as all charges accumulate at the wall for $w \le -2\al$. 

Interestingly, a similar third-order phase transition between the pushed and the pulled phase was recently found \cite{Cunden17a} by analysing large deviation functions associated with the position of the farthest charge in a $d$-dimensional jellium model. The limiting distribution of the position of the farthest charge is known
in $d=1$ (and was computed by Baxter, see Eq. (32) and (47)) and in $d=2$ where
the distribution, properly centred and scaled, approaches a Gumbel
distribution \cite{Rider14}. However, for $d>3$, no explicit result is known for
this limiting distribution.
In $d = 1$ this corresponds to the distribution of the maximum of $|x_i|$'s of the charges, as discussed above [see Eqs. (\ref{eq:ZNL}) and (\ref{Z_NL_prob})]. It was further shown that this observable exhibits a similar third order phase-transition even for short-range interactions, like the Yukawa potential \cite{Cunden17b}, in $d \geq 1$.  

\subsubsection{Right large deviation:}
We now focus on the distribution $Q(w,N)$, for large $N$, in the region $0<w-2\al \sim O(1)$, that characterises the large fluctuations of order $O(1)$ to the right of the mean. From the analysis performed in the previous section,  we have seen that in this regime, to leading order for large $N$, $Q(w,N) \approx 1$ [see  (\ref{Q_largew})]. To compute the sub leading corrections to this leading order term $1$, it is convenient to consider instead the PDF of $x_{\max}$, given by the derivative of  (\ref{Q_N-1}) 
\begin{align}
\label{exact_PDF}
P(w,N) =& \partial_w Q(w,N) = \frac{N}{Z_N(\infty)} \ee^{-\frac{N^2}{2}w^2} \int_{-\infty}^w dx_1 \cdots \int_{-\infty}^w dx_{N-1}\; \nonumber \\ 
 &\times  \exp \left(2\alpha\,N\, \sum_{j=1}^{N-1} |w-x_j| + \al\, N\,\sum_{1\leq i\neq j\leq N-1}|x_i-x_j| 
- \frac{N^2}{2}\sum_{i=1}^{N-1}x_i^2 \right)
\end{align}
where we have simply separated out the $x_N = w$ from the rest in Eqs. (\ref{Z_N-w}) and (\ref{E-1dc}). This can be re-written as
\begin{eqnarray}\label{PDF_largeN_1}
P(w,N) = \frac{N\,Z_{N-1}(\infty)}{Z_N(\infty)}\, \ee^{-\frac{N^2}{2}w^2} \Big \langle \ee^{2\al \,N\sum_{j=1}^{N-1}(w-x_j)}\Big\rangle_{N-1} \;,
\end{eqnarray}
where $\langle \cdots \rangle_{N-1}$ denotes the average over the Boltzmann distribution of $N-1$ charges. We can then analyse this average for large $N$, for $w > 2 \al$, following Ref. \cite{Majumdar09} for the log-gas in the corresponding right large deviation regime. To evaluate this average, we note that essentially one single charge out of $N$ is detached at $w > 2\al$, while the rest of $N-1$ charges should be in their equilibrium flat configuration, i.e., with a density $\rho^*_w(x) = 1/(4 \al)$ for $x \in [-2\al,2\al]$ (see the right panel of Fig. \ref{Fig_pushed_pulled}). Furthermore, for large $N$, to leading order, we can (i) approximate the average of the exponential in Eq.~(\ref{PDF_largeN_1}) by the exponential of the average and (ii) use that, to leading order for large $N$, $ \frac{N\,Z_{N-1}(\infty)}{Z_N(\infty)} \sim e^{-C_0\, N^2}$ for some constant $C_0$ (independent of $w$) to write
\begin{eqnarray}\label{PDF_largeN_2}
\hspace*{-2cm}P(w,N) \approx \ee^{-\frac{N^2}{2}w^2 + 2 \al \, N \langle \sum_{j=1}^{N-1} (w-x_j)\rangle-N^2\,C_0} \approx \ee^{-N^2\left(\frac{w^2}{2}-2\al \int_{-2\al}^{2\al} (w-x)\rho^*_w(x) dx - C_0\right)} \;.
\end{eqnarray}
Using $\rho^*_w(x) = 1/(4 \al)$ for $x \in [-2\al,2\al]$ and performing the integral in  (\ref{PDF_largeN_2}), we obtain 
\begin{eqnarray}\label{sm_Epulled}
P(w,N) \approx \ee^{-\Delta E_{\rm pulled}} \approx \ee^{-N^2\,  \Phi_+(w)} \;,
\end{eqnarray}
where
\begin{eqnarray}\label{sm-phi+}
\Phi_+(w) = \frac{(w-2\al)^2}{2} \;, \quad w > 2\al \;.
\end{eqnarray}
Thus $\Delta E_{\rm pulled}$ in  (\ref{sm_Epulled}) corresponds to the energy in pulling out a single charge from the equilibrium configuration of charges with a flat density.


\begin{figure}[t]
\centering
\includegraphics[scale=0.5]{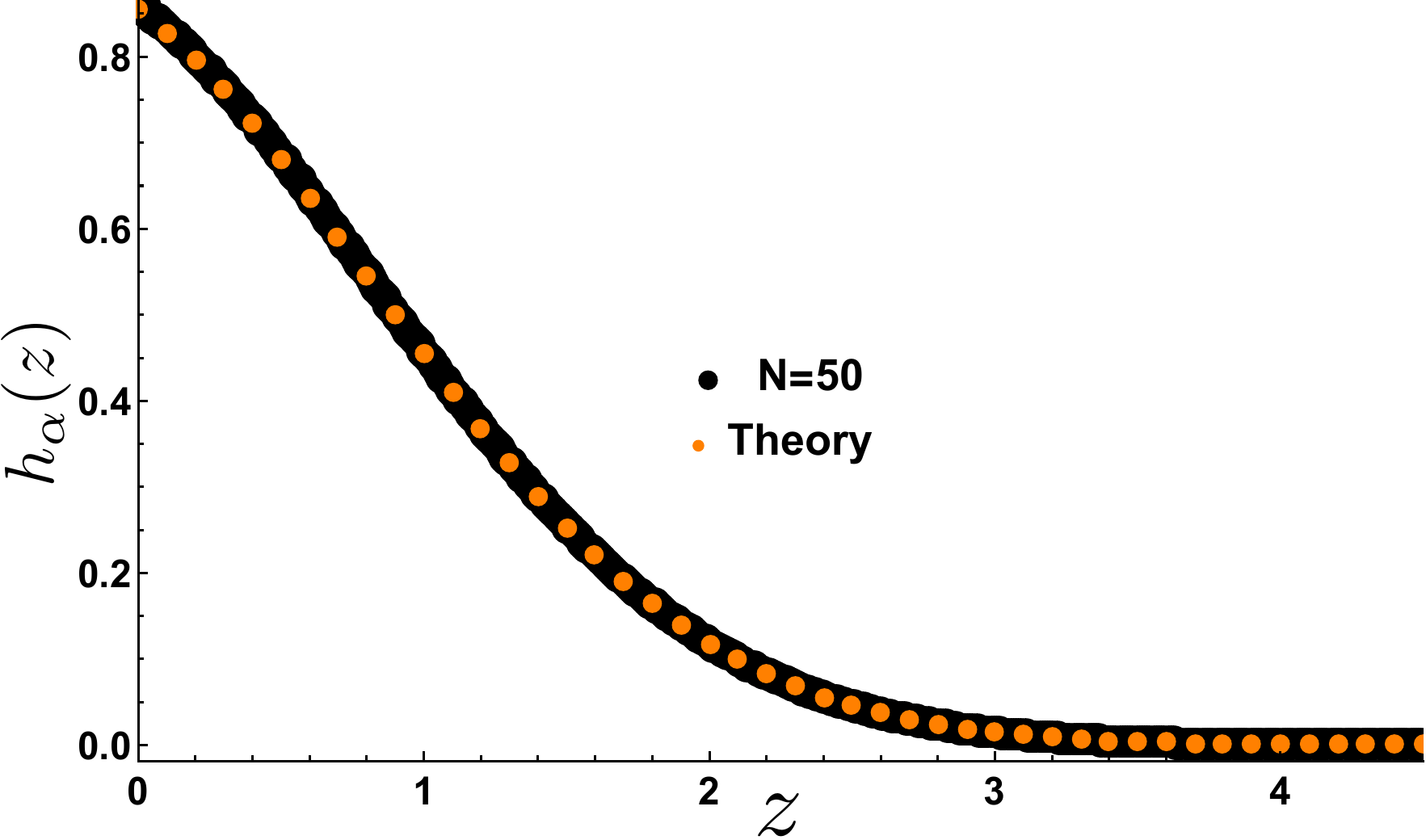} 
	\caption{\small Comparison of the gap distribution \eqref{h_a_5} with simulation. Simulations are performed for $N=50$ particles with $\alpha=0.1$.}  
	\label{fig:gap-dist}
\end{figure}

\section{Distribution of the gap $g=x_N-x_{N-1}$}
In this section we study the PDF $P_G(g,N)$ of the gap $g=x_N-x_{N-1}$ between the positions of the two rightmost charges with ordered positions $x_N$ and $x_{N-1}<x_N$. We show that the typical fluctuations of $g$, of order $O(1/N)$ for large $N$, are described by the scaling form,  
\begin{align}
P_G(g,N) &\approx N~h_\alpha(gN),~~\text{with}, \nonumber \\
h_\alpha(z) &= \Theta(z)\,A(\alpha)\,\int_{-\infty}^\infty dy~\e^{-u^2/2}~\frac{d}{du}F_\alpha(u-z+4\alpha)
\end{align}
where $F_\alpha(y)$ is given by the solution of \eqref{limit.1} with its associated eigenvalue $A(\alpha)$. For much larger values of $g$, i.e. $g = O(1)$, $P_G(g,N)$ is described by the large deviation form given in Eq. (\ref{large_dev_gap}).

The distribution of the gap $g=x_N-x_{N-1}$ can be formally written as 
\begin{equation}
P_G(g,N)= \frac{N!}{Z_N} \int dx_1\ldots dx_N \e^{-\beta E[\{x_i\}]} ~ \delta(x_N-x_{N-1}-g)~ \prod_{j=2}^N \Theta(x_j - x_{j-1}) \;,
\label{gap_dist-1}
\end{equation}
where $\beta E[\{x_i\}]$ is given in Eq. (\ref{partition_func}). In the ordered sector, rewriting the energy $E[\{x_i\}]$ as in \eqref{E-ordered} and performing the change of variables $x_i \to \ep_i$ as in \eqref{transformation}, one can write
\begin{equation}
P_G(g,N)=N^{1-N}\frac{N!}{Z_N} \int_{-\infty}^{\infty} \prod_{k=1}^N d\ep_k\, \e^{-\frac{1}{2}\sum_{k=1}^N \ep_k^2}\,
\prod_{k=2}^N \Theta(\ep_k-\ep_{k-1}+4\alpha)\, \delta(\ep_N-\ep_{N-1}-gN+4\alpha) \;.
\end{equation}
It is useful to regroup the pair of variables $\ep_N$ and $\ep_{N-1}$, keeping the rest $N-2$ variables together and rewrite the integral as
\begin{align}
&P_G(g,N)= N^{1-N}\frac{N!}{Z_N}\int_{-\infty}^{\infty} d\ep_N \int_{-\infty}^{\infty} d\ep_{N-1}\e^{-\frac{1}{2} (\ep_N^2+\ep_{N-1}^2)} \delta(\ep_N-\ep_{N-1}-gN+4\alpha)\, \nonumber \\ 
&~~~~~~~~~~~~~~~\times~\Theta(\ep_N-\ep_{N-1}+4\alpha)~\left(\int_{-\infty}^{\infty} \prod_{k=1}^{N-2} d\ep_k\, \e^{-\frac{1}{2}\sum_{k=1}^{N-2} \ep_k^2}\,
\prod_{k=2}^{N-1} \Theta(\ep_k-\ep_{k-1}+4\alpha)\right)\, \nonumber \\
\begin{split}
&~~= \frac{N!N^{1-N}}{Z_N}\int_{-\infty}^{\infty} d\ep_N \int_{-\infty}^{\infty} d\ep_{N-1}\e^{-\frac{1}{2} (\ep_N^2+\ep_{N-1}^2)} \delta(\ep_N-\ep_{N-1}-gN+4\alpha)\, \\ 
&~~~~~~~~~~~~~~~~~~~~~~~~~\times~\Theta(\ep_N-\ep_{N-1}+4\alpha)~D_\alpha(\ep_{N-1}+4\alpha,N-2)\;, \label{eq:PNg}  
\end{split}
\end{align}
where in the last step we have used the definition of $D_\alpha(x,N)$ in \eqref{D_N}. This formula for the PDF of the gap $P_G(g,N)$ in Eq. (\ref{eq:PNg}) is exact for any $N$. We now analyze it in the large $N$ limit. In this limit, it turns out that the typical fluctuations of the gap are of order $O(1/N)$, as suggested by the appearance of the scaling variable $g\,N$ in Eq.~(\ref{eq:PNg}), while, as for $x_{\max}$, the atypical fluctuations are of order $O(1)$. We now analyse separately these two regimes of typical and atypical fluctuations of the first gap.

\subsection{Typical fluctuations of the gap $g$}

To analyse the typical fluctuations of the gap $g$, we consider the limit $N \to \infty$, $g \to 0$, keeping the scaling variable $z = N\,g$ fixed. For large we use Eq. (\ref{def_FN}) to write $D_\alpha(x,N-2)= D_\alpha(\infty,N-2) F_\alpha(x,N-2)$. Using further
$Z_N \approx N!D_\alpha(\infty,N)/N^N$, we have
 \begin{align}
P_G(g,N) \approx & N\frac{D_\alpha(\infty,N-2)}{D_\alpha(\infty,N)}\int_{-\infty}^{\infty} d\ep_N \int_{-\infty}^{\infty} d\ep_{N-1}\e^{-\frac{1}{2} (\ep_N^2+\ep_{N-1}^2)} \delta(\ep_N-\ep_{N-1}-gN+4\alpha) \nonumber \\
&~~~~~~~~~~~~~~~~~~~~~~~\times~~\Theta(\ep_N-\ep_{N-1}+4\alpha)\,~F_\alpha(\ep_{N-1}+4\alpha,N-2). 
\end{align}
In the large $N$ limit, we then use the fact that $D(\alpha,N)\sim [A(\alpha)]^{-N}$ and that $F_\alpha(x,N \to \infty) = F_\alpha(x)$. Further, keeping $N\,g  =z$ fixed in the scaling limit, we get 
 \begin{equation}
P_G(g,N) \approx NA(\alpha)^2\int_{-\infty}^{\infty} dx \int_{-\infty}^{\infty} dy~\e^{-\frac{1}{2} (x^2+y^2)} \delta(x-y-gN+4\alpha) 
~\Theta(x-y+4\alpha)\,~F_\alpha(y+4\alpha), \label{PN_g}
\end{equation}
where $F_\alpha(x)$ satisfies the differential equation \eqref{limit.1}. Clearly, $P_G(g,N)$ in Eq. (\ref{PN_g}) has the scaling form 
\begin{eqnarray}\label{scaling_PNg1}
P_G(g,N) &\approx& N\,h_\alpha(g\,N) \;, 
\end{eqnarray}
where the scaling function $h_\alpha(z)$ is given by
\begin{eqnarray}\label{h_a_1}
\hspace*{-2.5cm}h_\alpha(z) = A(\alpha)^2\int_{-\infty}^{\infty} dx \int_{-\infty}^{\infty} dy~\e^{-\frac{1}{2} (x^2+y^2)} \delta(x-y-z+4\alpha) ~\Theta(x-y+4\alpha)\,~F_\alpha(y+4\alpha)\,. \nonumber \\
\end{eqnarray}
This scaling function $h_\alpha(z)$ in Eq. (\ref{h_a_1}) can be further simplified as follows. The presence of the theta function in Eq. (\ref{h_a_1}) indicates that this integral is non-zero only when $x > y - 4 \alpha$. In contrast, the delta function indicates that this is non-zero only when $x = y+z-4\al$. Hence, for $z<0$, these two conditions can not be satisfied simultaneously. This indicates that $h_\alpha(z<0) = 0$. For $z>0$, the once the delta function constraint is satisfied, then the theta function constraint is automatically satisfied. Hence, for $z>0$, we can write
\begin{eqnarray}\label{h_a_2}
\hspace*{-1.5cm}h_\alpha(z) = \Theta(z)\,A^2(\alpha)\,  \int_{-\infty}^{\infty} dx \int_{-\infty}^{\infty} dy~\e^{-\frac{1}{2} (x^2+y^2)} \delta(x-y-z+4\alpha)~F_\alpha(y+4\alpha) \;.
\end{eqnarray}
Performing the integral over $x$, we get
\begin{eqnarray}\label{h_a_3}
\hspace*{-0.cm}h_\alpha(z) = A^2(\alpha)\, \Theta(z) \int_{-\infty}^\infty dy\, \e^{-\frac{y^2}{2}} \e^{-\frac{(y+z-4\al)^2}{2}}~F_\alpha(y+4\alpha) \;.
\end{eqnarray}
Using the differential equation (\ref{limit.1}) one can simplify further 
\begin{eqnarray}\label{h_a_4}
\hspace*{-0.cm}h_\alpha(z) = A(\alpha) \Theta(z) \int_{-\infty}^\infty dy\, \e^{-\frac{1}{2}(y+z-4\al)^2} F_\alpha'(y) \;.
\end{eqnarray}
One can also do an integration by parts to rewrite it as
\begin{eqnarray}\label{h_a_5}
\hspace*{-0.cm}h_\alpha(z) = A(\alpha) \Theta(z) \int_{-\infty}^\infty dy \, (y+z-4\al) \e^{-\frac{1}{2}(y+z-4\al)^2} F_\alpha(y) \;.
\end{eqnarray}
In Fig. \ref{fig:gap-dist} we compare this theoretical result with numerical simulation and observe a very good agreement. One can also
estimate the asymptotic tails of the scaling function $h_\alpha(z)$. From Eq. (\ref{h_a_5}), as $z \to 0$, the scaling function $h_\alpha(z)$ approaches a constant given~by
\begin{eqnarray}\label{h0}
h_\alpha(0) = A_\alpha \int_{-\infty}^\infty dy \, (y-4\al) \e^{-\frac{1}{2}(y-4\al)^2} F_\alpha(y) \;,
\end{eqnarray}
which can be evaluated numerically. For $z \to \infty$, one can show, by analysing the integral in Eq. (\ref{h_a_5}) and using the tails of $F_\alpha(y)$ from Eq. (\ref{fx_asymp}), that to leading order for large $z$, 
\begin{eqnarray}\label{h_largez}
h_\alpha(z) \sim \e^{-z^2/2 + o(z^2)} \;.
\end{eqnarray}
As expected, this is similar to the right tail behaviour of the PDF of $x_N = x_{\max}$ [see Eq.~(\ref{fx_asymp})], since to create a large gap, we must have $x_N \gg x_{N-1}$. The fact that the large gap asymptotic behaviour coincides with the right tail of $x_{\max}$ also holds for the log-gas case \cite{Perret14}.

\subsection{Atypical large fluctuations of the gap $g$}

To analyse the large fluctuations of the gap $g$ of order $O(1)$, it is useful to remark that the configurations
that contribute to the PDF $P(g,N)$ are the same that contribute to a large value of $x_{\max}$ to the right of
its mean, as depicted in the right panel of Fig. \ref{Fig_pushed_pulled}. In such configurations, $x_N = x_{\max}  = w$
while the second particle $x_{N-1}$ is located close to the right edge $x_{N-1} \approx 2 \alpha$, leading to a gap $g = w-2\alpha$. Therefore, for large $N$, one obtains that ${\rm Prob.}(x_{N}-x_{N-1} = g) \approx {\rm Prob.}(x_{\max} = g + 2 \al)$. Therefore, from the right large deviation form of the PDF of $x_{\max}$ obtained in Eqs. (\ref{sm_Epulled}) and (\ref{sm-phi+}), one gets, the large deviation form of $P(g,N)$ to leading order for large $N$ as
\begin{eqnarray}\label{large_dev_gap}
P(g,N) \sim \e^{-N^2 \Psi_+(g)} \;, \; \Psi_+(g) = \Phi_+(g + 2\al) = \frac{g^2}{2} \;,
\end{eqnarray} 
where we have used the explicit expression of $\Phi_+(w)$ given in Eq. (\ref{sm-phi+}). Interestingly, this large deviation regime coincides exactly with the right tail of the regime of typical fluctuations [see Eq. (\ref{scaling_PNg1}) and (\ref{h_largez})].

\section{Index distribution}
\label{index}
In this section, we study the statistics of the index $N_+$, which is the number of charges located on the positive semi-axis, 
i.e. $N_+=\sum_{i=1}^N \Theta(x_i)$. Clearly, $N_+$ is a random variable with range $0\le N_+\le N$ and we now compute its distribution $P_I(N_+,N)$ for large $N$. It is also clear that $\langle N_+\rangle=N/2$ and the distribution $P_I(N_+,N)$ is symmetric around this mean. Given the joint PDF $\mathcal{P}(x_1,x_2,\ldots,x_N)$ in \eqref{jpdf-1dc} along with~\eqref{E-1dc_2}, $P_I(N_+,N)$ can be expressed as a multiple integral
\begin{equation}
P_I(N_+,N)= \int_{-\infty}^{\infty}dx_1\dots \int_{-\infty}^{\infty}dx_N\,  \mathcal{P}(x_1,x_2,\dots,x_N)\, \delta\left[
\sum_{i=1}^N \Theta(x_i)-N_+\right].
\label{index_dist.1}
\end{equation}
Since the integrand in  (\ref{index_dist.1}) is symmetric under any permutation of the $x_i$'s, we can
order the $x_i$'s, with $x_1<x_2< x_3\dots < x_N$ and rewrite it as
\begin{equation}
P_I(N+,N)= \frac{N!}{Z_N} \int dx_1\ldots dx_N \ee^{-\beta\,E[\{x_i\}]} \delta\left[
\sum_{i=1}^N \Theta(x_i)-N_+\right]\, \prod_{j=2}^N \Theta(x_j - x_{j-1}) 
\label{index_dist.2} \;,
\end{equation}
where $\beta\,E[\{x_i\}]$ is given in Eq. (\ref{E-1dc_2}). In this ordered sector $(x_1<x_2<x_3\ldots<x_N)$, we use the same trick to eliminate the absolute values as done in \eqref{E-ordered}.
Hence, up to an overall normalisation constant, we can write
\begin{align}
 \begin{split}
P_I(N_+,N) \propto & \int_{-\infty}^{\infty}dx_1\dots \int_{-\infty}^{\infty} dx_N \,
\e^{-\frac{N^2}{2}\sum_{i=1}^N [x_k- \frac{2\alpha}{N} (2k-N-1)]^2}\,  \\ 
&~~~~~~~~\times~\prod_{j=2}^N \Theta(x_j - x_{j-1}) \, \delta\left[
\sum_{i=1}^N \Theta(x_i)-N_+\right]\, .
\end{split}
\label{index_dist.3}
\end{align} 

\begin{figure}[t]
		\centering
		\includegraphics[scale=0.5]{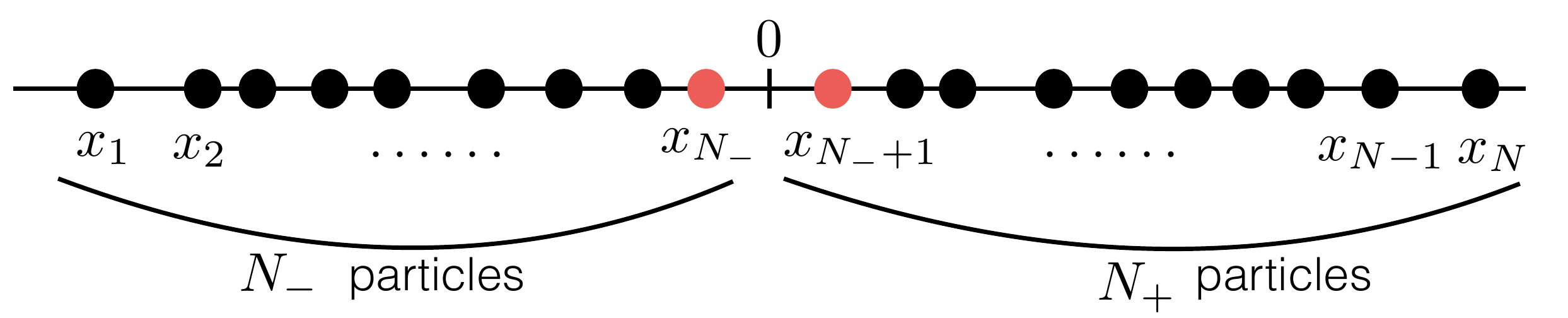} 
	\caption{\small A typical configuration of the positions of the charges having $N_-$ particles on the negative axis and $N_+$ charges on the positive axis.}  
	\label{fig:index-config}
\end{figure}

\noindent
Next we perform the change of variables given in \eqref{transformation} and rewrite the product of theta functions
in Eq. (\ref{index_dist.3}) as
\begin{equation}
\prod_{k=2}^N \Theta(x_k-x_{k-1})= \prod_{k=2}^N \Theta(\ep_k-\ep_{k-1}+4\alpha)\, .
\label{constraint.1}
\end{equation}
We want to compute the probability that $N_+$ charges are on the positive side, or equivalently that $N_{-}=N-N_+$ charges are on the negative side.
Consider first the $N_{-}$ charges on the negative side. In the ordered sector, we have to ensure
that the position of the $N_{-}$'th charge is negative. This automatically ensures (since we are in the ordered 
sector) that all the $N_{-}$ charges with positions $x_1$, $x_2$, $\dots$, $x_{N_{-}}$ are negative. Thus, using the variables $\epsilon_i$'s in Eq. (\ref{transformation}), this condition translates to (see Fig. \ref{fig:index-config})
\begin{equation}
x_{N_{-}}<0 \quad\quad {\rm or\,\, equivalently}\quad\quad \ep_{N_{-}}< 2\alpha(N+1-2N_{-})= 
4\alpha \left(N_+-\frac{N}{2}\right)+ 2\alpha
\label{cond.1}
\end{equation}
where we have used  (\ref{transformation}) with $i=N_{-}$ and $N_{-}=N-N_+$. Similarly, the condition $x_{N_{-}+1}>0$ automatically 
ensures (in the ordered sector) that the position of all $N_+$ charges on the right are positive, i.e.,
$x_{N_{-}+1}>0$, $x_{N_{-}+2}>0$, $\dots$, $x_N>0$. Thus this condition translates to (see Fig. \ref{fig:index-config})
\begin{equation}
x_{N_{-}+1}>0 \quad\quad {\rm or\,\, equivalently}\quad\quad \ep_{N_{-}+1}> 2\alpha(N-1-2N_{-})
= 4\alpha \left(N_+-\frac{N}{2}\right)- 2\alpha\, .
\label{cond.2}
\end{equation} 
For later convenience, let us define 
\begin{equation}
z= 4\alpha \left(N_+- \frac{N}{2}\right)\, .
\label{def_z}
\end{equation}
Then, in terms of the $z$ variable, the two conditions in  (\ref{cond.1}) and  (\ref{cond.2}) are expressed as
\begin{equation}
\ep_{N_{-}}< z+ 2\alpha \quad \quad {\rm and}\quad\quad \ep_{N_{-}+1}> z- 2\alpha \, .
\label{cond.3}
\end{equation}
Thus finally, using these new variables, $P_I(N_+,N)$ in  (\ref{index_dist.3}) simplifies to
\begin{align}
\begin{split}
&P_I(N_+,N) \propto T(N_+,N),~~~~\text{where}, \\ 
&T(N_+,N)= \int_{-\infty}^{\infty} \prod_{k=1}^N d\ep_k\, \ee^{-\frac{1}{2}\sum_{k=1}^N \ep_k^2}\,
\prod_{k=2}^N \Theta(\ep_k-\ep_{k-1}+4\alpha)\, \\ 
&~~~~~~~~~~~~~~~~~~~~~~~~~~~~~
\times~ \Theta(z+2\alpha- \ep_{N_{-}})\, \Theta(\ep_{N_{-}+1}- z+2\alpha)\, ,
\end{split}
\label{index_dist.4}
\end{align}
where we have used the condition in  (\ref{constraint.1}), as well as the two conditions in  (\ref{cond.3}).
Note that these equations are strictly valid for $0<N_+<N$. For $N_+=N$ and  $N_+=0$, one has to 
consider a slightly different integral, but this does not make any difference in the scaling limit.
Thus, once again we have reduced our original problem of a long-ranged Coulomb gas to a problem of a short-ranged gas where
there are only nearest neighbour interactions. Additionally, now there is a `defect' on the bond connecting $\ep_{N_{-}}$
and its right neighbour $\ep_{N_{-}+1}$ that makes this short-ranged gas inhomogeneous (see Fig.~\ref{fig:index-config}).

The integral in  (\ref{index_dist.4}) can be further simplified into two blocks as follows.
For simplicity, let us denote the value of $\ep$'s across the `defect' as
\begin{equation}
\ep_{N_{-}}=u \quad\quad {\rm and} \quad\quad \ep_{N_{-}+1}=v\, .
\label{def_uv}
\end{equation}
Then the integral in  (\ref{index_dist.4}) can be expressed as
\begin{equation}
T(N_+,N) = \int_{-\infty}^{\infty} du \int_{-\infty}^{\infty} dv \, T_1(u)\, T_2(v)\, \ee^{-\frac{1}{2} (u^2+v^2)}\, 
\Theta(v-u+4\alpha)\, \Theta(z+2\alpha- u)\, \Theta(v-z+2\alpha)\, ,
\label{index_dist.5}
\end{equation}
where $T_1(u)$ is the integral over the left $M=N_{-}-1$ variables (given $\ep_{N_{-}}=u$) and $T_2(v)$
is the integral over the right $N_{+}-1=N-M-2$ variables (given $\ep_{N_{-}+1}=v$). They are given explicitly by
\begin{equation}
T_1(u)= \int_{-\infty}^{\infty}d\ep_1\int_{-\infty}^{\infty} d\ep_2\dots \int_{-\infty}^{\infty} d\ep_{M}\,
\ee^{-\frac{1}{2} \sum_{k=1}^{M} \ep_k^2}\, \left[\prod_{k=2}^{M}\Theta(\ep_k-\ep_{k-1}+4\alpha)\right]\, 
\Theta(u-\ep_M+4\alpha)\, 
\label{T1u}
\end{equation}
where we recall that $M= N_{-}-1$. Similarly, for the right block, we have
\begin{align}
\begin{split}
T_2(v)= & \int_{-\infty}^{\infty}d\ep_{M+3}\int_{-\infty}^{\infty} d\ep_{M+4}\dots 
\int_{-\infty}^{\infty} d\ep_{N}\,~~
\ee^{-\frac{1}{2} \sum_{k=M+3}^{N} \ep_k^2}\, \\ 
&~~~~~~~~~~~~~~~~~~~~~~~~~~~
\times ~ \left[\prod_{k=M+4}^{N}\Theta(\ep_k-\ep_{k-1}+4\alpha)\right]\,
\Theta(\ep_{M+3}-v+4\alpha)\,.
\end{split}
\label{T2v}
\end{align}
We can re-write $T_1(u)$ in  (\ref{T1u}) by incorporating the constraints imposed by the
theta functions directly in the limits of integration as
\begin{equation}
T_1(u)= \int_{-\infty}^{u+4\alpha} d\ep_M\, \ee^{-\frac{1}{2}\ep_M^2}\,\int_{-\infty}^{\ep_M+4\alpha} d\ep_{M-1} 
\, \ee^{-\frac{1}{2}\ep_{M-1}^2}\ldots \int_{-\infty}^{\ep_2+4\alpha} d\ep_1\, 
\ee^{-\frac{1}{2}\ep_{1}^2}
\label{T1u.1}
\end{equation}
where we recall again $M=N_{-}-1$. Note that this is exactly, the function $D_\alpha(x,N)$ defined in \eqref{D_N}. 
Hence, $T_1(u)$ in (\ref{T1u.1}) reads
\begin{equation}
T_1(u)= D_\alpha(u+4\alpha, M)\, , \quad\quad {\rm where}\quad M= N_{-} - 1= N-N_+-1\;.
\label{T1u.2}
\end{equation}
In a similar way, we can rewrite the integral $T_2(v)$ in  (\ref{T2v}) as
\begin{equation}
T_2(v)= \int_{v-4\alpha}^{\infty} d\epsilon_{N_{-}+2}\, \ee^{-\frac{1}{2}\epsilon_{N_{-}+2}^2}\,
\int_{\epsilon_{N_{-}+2}-4\alpha}^{\infty} d\epsilon_{N_{-}+3}\,  \ee^{-\frac{1}{2}\epsilon_{N_{-}+3}^2}\ldots
\int_{\epsilon_{N-1}-4\alpha}^{\infty} d\epsilon_N\, \ee^{-\frac{1}{2}\epsilon_N^2}\, .
\label{T2v.1}
\end{equation}
As in the case of the left block, let us define a function similar to $D_\alpha(x,M)$ 
\begin{equation}
E_\alpha(x, M)= \int_{x}^{\infty} dy_1\, \ee^{-\frac{1}{2} y_1^2}\, \int_{y_1-4\alpha}^{\infty} dy_2\, 
\e^{-\frac{1}{2} y_2^2} \ldots \int_{y_{M-1}-4\alpha}^{\infty}dy_M\, \ee^{-\frac{1}{2} y_M^2}\, .
\label{Ex_def}
\end{equation}
Then, $T_2(v)$ in  (\ref{T2v.1}) can be written simply as
\begin{equation}
T_2(v)= E_\alpha(v-4\alpha, N_+ -1)\, .
\label{T2v.2}
\end{equation}
Finally, from the definitions of $D_\alpha(x,M)$ (\ref{D_N}) and $E_\alpha(x,M)$ (\ref{Ex_def}), it is easy to check, by performing 
the change of variables $y_k\to -y_k$, the following identity
\begin{equation}
D_\alpha(x,M)= E_\alpha(-x,M)
\label{DE.0}
\end{equation}
valid for all $M\ge 0$. Plugging the results from Eqs. (\ref{T1u.2}) and (\ref{T2v.2}) into  (\ref{index_dist.5}) gives 
\begin{eqnarray}
\hspace*{-2.cm}T(N_+,N) &=& \int_{-\infty}^{z+2\alpha} du\, \int_{z-2\alpha}^{\infty} dv \,\, 
D_\alpha(u+4\alpha, N-N_+ -1)\,\, E_\alpha(v-4\alpha, N_+ -1)\,\, \nonumber \\
&\times& \ee^{-\frac{1}{2} (u^2+v^2)}\,
\,\Theta(v-u+4\alpha)\, 
\label{index_dist.6}
\end{eqnarray}
where $z= 4\alpha \left(N_+-\frac{N}{2}\right)\,$ from  (\ref{def_z}).
The double integral in  (\ref{index_dist.6}) can be further simplified
by making the following observation. Let us look at the range of integration of $u$. The theta function
$\Theta(v-u+4\alpha)$ demands that $u<v+4\alpha$. Hence we can eliminate the theta function and write it as
\begin{equation}
\small
T(N_+,N) = \int_{z-2\alpha}^{\infty} dv \,\, \int_{-\infty}^{\min (v+4\alpha, z+2\alpha)} du\,
D_\alpha(u+4\alpha, N-N_+ -1)\,\, E_\alpha(v-4\alpha, N_+ -1)\,\, \ee^{-\frac{1}{2} (u^2+v^2)}\, .
\label{index_dist.7}
\end{equation} 
However, the lower limit of the $v$ integration implies $v>z-2\alpha$. This means
$v+4\alpha>z+2\alpha$. Hence, we necessarily have, ${\min (v+4\alpha, z+2\alpha)}= z+2\alpha$.
Thus we get
\begin{equation}
T(N_+,N) = \int_{z-2\alpha}^{\infty} dv \,\, \int_{-\infty}^{z+2\alpha}du\,
D_\alpha(u+4\alpha, N-N_+ -1)\,\, E_\alpha(v-4\alpha, N_+ -1)\,\, \ee^{-\frac{1}{2} (u^2+v^2)}\, .
\label{index_dist.8}
\end{equation}
Making further the change of variable $v\to -v$, we can write it as
\begin{equation}
T(N_+,N) = \int_{-\infty}^{-z+2\alpha} dv \,\, \int_{-\infty}^{z+2\alpha} du\,
D_\alpha(u+4\alpha, N-N_+ -1)\,\, E_\alpha(-v-4\alpha, N_+ -1)\,\, \ee^{-\frac{1}{2} (u^2+v^2)}\, .
\label{index_dist.9}
\end{equation}
Finally, using $E_\alpha(-x,M)= D_\alpha(x,M)$ from  (\ref{DE.0}), we get 
\begin{align}
\begin{split}
&P_I(N_+,N)\propto T(N_+,N),~~~\text{with}\\
&T(N_+,N)=\left[\int_{-\infty}^{z+2\alpha} D_\alpha(u+4\alpha, N-N_+-1)\, \ee^{-\frac{1}{2} u^2}\, du\right] \\ 
&~~~~~~~~~~~~~~~~~~~~~~~~~~
\times~\left[\int_{-\infty}^{-z+2\alpha}  D_\alpha(v+4\alpha, N_+-1)\, \ee^{-\frac{1}{2} v^2}\, dv\right] 
\end{split}
\label{index_dist.10}
\end{align}
where we recall $z= 4\alpha \left(N_+- \frac{N}{2}\right)$.
Note that the rhs of  (\ref{index_dist.10}) 
is manifestly a symmetric function of $N_+$ around $N_+=N/2$.

This result in  (\ref{index_dist.10}) is exact for all $1<N_+<N$. We now analyse it in the large $N$ limit. For large $N$, it turns out that the typical fluctuations of $N_+$ around its mean $N_+=N/2$ are of order $O(1)$ (i.e. $z = O(1)$), while the atypical fluctuations are of order $O(N)$ (i.e. $z = O(N)$). Below, we analyse separately the probability distribution of typical and atypical fluctuations.

\subsection{Typical fluctuations of $N_+$}

It is convenient to define the fraction $c= N_+/N$ with $0\le c\le 1$. In the typical regime, where $N_+ = N/2 + z/(4\al)$ with $z = O(1)$.  This amounts to consider the scaling limit $c\to 1/2$, $N\to \infty$, while keeping the product
$z= 4\alpha N(c-1/2)$ finite,  $O(1)$.  
To analyse  (\ref{index_dist.10}) in this scaling limit ($N\to \infty$ with
$z$ fixed), we follow the method of Section \ref{P_x_max} and rewrite the function $D_\alpha(x,M)$ as 
\begin{equation}
F_\alpha(x,M)= \frac{D_\alpha(x,M)}{D_\alpha(\infty,M)}. 
\label{Fx_def}
\end{equation}
In Section \ref{P_x_max} ,we have  shown that in the large $M$ limit, $D_\alpha(\infty, M) \sim [A(\alpha)]^{-M}$
where $A(\alpha)$ can be interpreted as the free energy associated to the short-ranged gas whose partition function is given in Eq. (\ref{D_N}). Furthermore, the function $F_\alpha(x,M)$ converges to a $M$ independent limiting function $F_\alpha(x)$ [as stated in \eqref{limit.0}] where $F_\alpha(x)$ satisfies the nonlocal eigenvalue equation \eqref{limit.1}.
We then replace $D_\alpha(x,M)= D_\alpha(\infty,M) F_\alpha(x,M)$ in  (\ref{index_dist.10}),
take the scaling limit using \eqref{D-inf} and \eqref{limit.0} and obtain
\begin{equation}
P_I(N_+,N) \propto    
\left[\int_{-\infty}^{z+2\alpha} A(\alpha)~F_\alpha(u+4\alpha)\, \ee^{-\frac{1}{2} u^2}\, du\right]
\left[\int_{-\infty}^{-z+2\alpha}  A(\alpha)~F_\alpha(v+4\alpha)\, \ee^{-\frac{1}{2} v^2}\, dv\right]\,,
\label{index_dist.11}
\end{equation}
where we have absorbed the prefactor $A(\al)^{-N}$ in the proportionality constant. Furthermore, by using \eqref{limit.1}, the integrals over $u$ and $v$ can be performed explicitly [using $F_\alpha(x \to -\infty) = 0$, see Eq. (\ref{Falpha-inf})]. 
This gives
\begin{equation}
P_I(N_+,N) \propto F_\alpha(z+2\alpha)\, F_\alpha(-z+2\alpha)\, .
\label{index_result.1}
\end{equation}
The proportionality constant can be fixed using the overall normalisation $\sum_{N_+=0}^N P_I(N_+,N)=1$. 

\begin{figure}[t]
		\centering
		\includegraphics[scale=0.5]{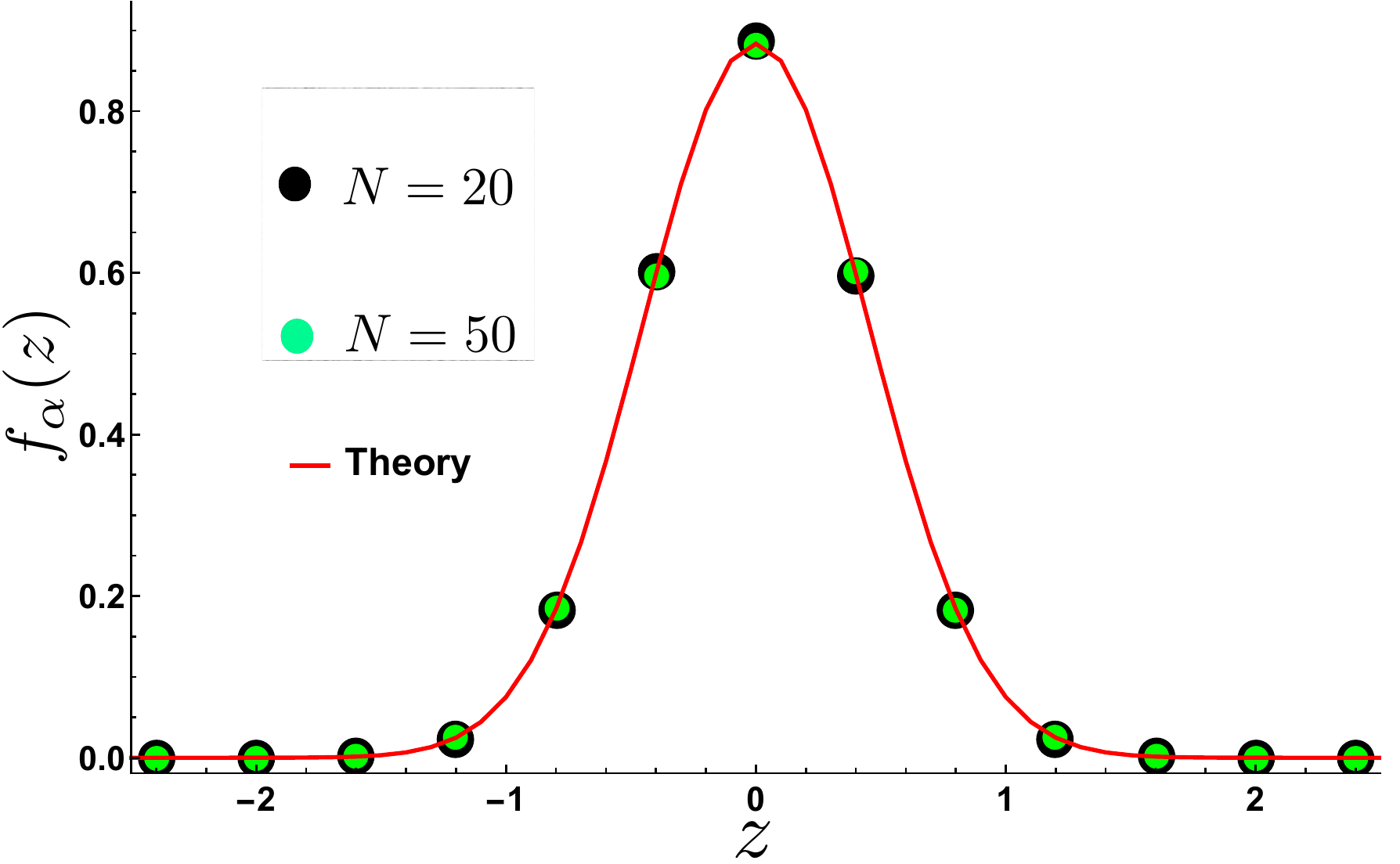} 
	\caption{\small Comparison of the limiting index distribution \eqref{limiting.2} with simulation. Simulations are performed for $N=20$ and $50$ particles with $\alpha=0.1$. }  
	\label{index-dist}
\end{figure}

Summarising, the random variable $N_+$ typically fluctuates on a scale of $O(1)$ around its mean value $N/2$. We find that as $N\to \infty$
\begin{equation}
N_+ \to \frac{N}{2} + \frac{1}{4\alpha}\, z
\label{typical_scaling}
\end{equation}
where the random variable $z$ has a limiting $N$-independent distribution $f_\alpha(z)$. 
In other words,
the distribution $P_I(N_+,N)$ converges to a limiting
scaling form in the large $N$ limit
\begin{equation}
P_I(N_+, N) \to 4\alpha\, f_\alpha\left( 4 \alpha \left(N_+-\frac{N}{2}\right)\right)\, ,
\label{limiting.1}
\end{equation}
where the scaling function $f_\alpha(z)$ is given from (\ref{index_result.1}) as
\begin{equation}
f_\alpha(z)=  \frac{F_\alpha(z+2\alpha)\, F_\alpha(-z+2\alpha)}{\int_{-\infty}^{\infty} dz\, 
F_\alpha(z+2\alpha)\, F_\alpha(-z+2\alpha)}.
\label{limiting.2}
\end{equation}
The function $f_\alpha(z)$ is manifestly symmetric around $z=0$. In Fig. \ref{index-dist} we compare this analytical result with numerical simulations and observe excellent agreement.  
The asymptotic behaviour of $f_\alpha(z)$ for large $z$ can be easily derived
using the asymptotic decay of $F_\alpha(z\to -\infty) \sim \exp[- |z|^3/{24\alpha}]$ [see Eq. \eqref{fx_asymp}] and the fact that $F_\alpha(z\to \infty)=1$. Plugging these asymptotics in~(\ref{limiting.2}) gives
\begin{equation}
f_\alpha(z) \sim \exp[- \frac{1}{24\alpha}\,|z|^3]\quad {\rm as}\quad |z|\to \infty \, .
\label{asymp.1}
\end{equation}
Thus the limiting distribution $f_\alpha(z)$ in Eq. (\ref{limiting.2}) is obviously non-Gaussian. This is at variance with the log-gas where the typical fluctuations of the index are known to be Gaussian [see Eq. (\ref{gauss_index})]. Furthermore, as we will see below, this tail behaviour from the central regime matches smoothly with the large deviation behavior of $N_+$.


\subsection{Atypical large fluctuations of the index $N_+$}
\label{LDF-index}
In this section we study large deviations regime of $P_I(N_+,N)$ in \eqref{index_dist.1} where $N_+ - N/2 = O(N)$ in the large $N$ limit. Our starting point is the exact expression for $P_I(N_+,N)$ in \eqref{index_dist.2} which we write as
\begin{align}
\begin{split}
P_I(N_+,N)&=\frac{N!~I(N_+,N)}{Z_N} \;, \\
I(N_+,N)&=\int dx_1\ldots dx_N \e^{-\beta\, E[\{x_i\}]} \delta\left[
\sum_{i=1}^N \Theta(x_i)-N_+\right]\,,\\ 
\end{split}
\label{I_N+}
\end{align}
with  $\beta \,E[\{x_i\}]$ given in \eqref{E-1dc_2}. Hence, $I(N_+,N)$ can be interpreted as 
the partition function of the $1d$ jelllium under the external constraint that there are exactly $N_+$ particles on the positive side. As the function $I(N_+=cN,N)$ is symmetric around $c=1/2$, we assume, for convenience, that $0\leq c \leq 1/2$.

 To proceed further, we follow the same Coulomb gas method as explained in Section~\ref{LDF-xmax}, to compute the partition function $I(N_+=cN,N)$. We first replace the multiple integrals over $x_i$'s in Eq. (\ref{I_N+}) by a functional integral over possible densities  
\begin{equation}
\rho_I(x,N)=N^{-1}\sum_i\delta(x-x_i) \;, \label{den-1}
\end{equation} 
where the subscript $I$ refers to ``index''. The density $\rho_I$ (i) is normalised and (ii) satisfies the constraint that $N_+ = c\,N$ charges are on the positive axis, i.e.
\begin{eqnarray}\label{constraint_rho_I}
(i) \;\; \int_{-\infty}^\infty dx \, \rho_I(x,N) = 1 \;, \;\;\; (ii) \int_{-\infty}^\infty dx \, \Theta(x) \rho_I(x,N) = c \;.
\end{eqnarray}
Therefore, the partition function $I(cN,N)$ reads, to leading order for large $N$ 
 \begin{equation}
I(cN,N) \propto \int \mathcal{D}[\rho_I]~e^{-N^3\Sigma_c[\rho_I]} \;,
\label{I_c}
\end{equation}
with
\begin{align}
\Sigma_c[\rho_I]=&~\frac{1}{2} \int_{-\infty}^{\infty} dx~x^2\rho_I(x) - \alpha \int_{-\infty}^{\infty} dx \int_{-\infty}^{\infty} dy~\rho_I(x)\rho_I(y)~|x-y| \nonumber \\ 
&~~~~~+ A_1\left( \int_{-\infty}^{\infty}dx~\rho_I(x) - 1\right) + A_2\left( \int_{-\infty}^{\infty}dx~\Theta(x)\rho_I(x) - c\right), \label{Sigma-1}
\end{align}
where $A_1$ and $A_2$ are Lagrange multipliers to enforce the constraints satisfied by $\rho_I(x)$~(\ref{constraint_rho_I}). 

\begin{figure}[t]
		\centering\includegraphics[scale=0.5]{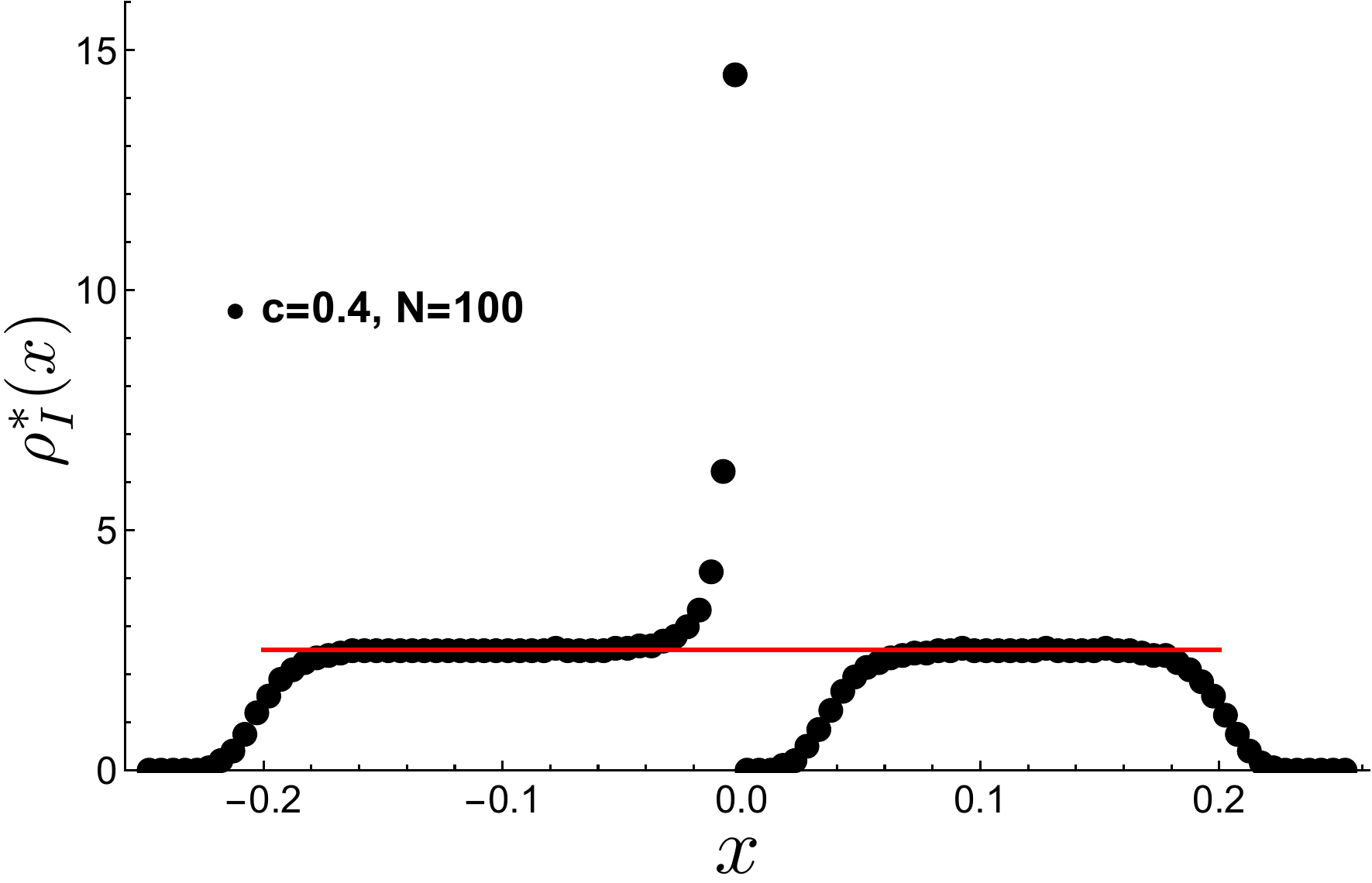} 
	\caption{\small Density profile associated to $c=0.4$ obtained from simulation with $N=100$. The red  horizontal line correspond to the bulk density $\rho(x)=1/(4 \alpha)$ in the unconstrained gas. }  
	\label{den-constrained}
\end{figure}

In the large $N$ limit, the functional integral in Eq. (\ref{I_c}) is dominated by the charge density $\rho_I^*$ that minimise $\Sigma_c[\rho_I]$. Numerical simulations indicate [see Fig. \ref{den-constrained}] that $\rho_I^*$ is of the form
\begin{eqnarray}
\rho_I^*(x) = 
\begin{cases}
&\rho_1(x) + \lambda~\delta(x),~~\text{for}~~-B\leq x \leq 0 \\
&\rho_2(x)~~~~~~~~~~~~~~~\text{for}~~~~~~~a\leq x \leq b 
\end{cases},
\label{den-anticipate}
\end{eqnarray}
where the constants $\lambda >0,~B>0,~a>0$ and $b>0$ have to be determined. Inserting this form of density in 
the functional $\Sigma_c[\rho]$ in \eqref{Sigma-1} we obtain 
\begin{align}
\begin{split}
\Sigma_c[\rho^*_I]= &~\frac{1}{2} \int_{-B}^{0} dx~x^2\rho_1(x)+\frac{1}{2} \int_{a}^{b} dx~x^2\rho_2(x) \\
&~~~- \alpha \left( \int_{-B}^{0} dx\int_{-B}^{0} dy~\rho_1(x)\rho_1(y)|x-y| + 2 \lambda \int_{-B}^{0} dx~|x|~\rho_1(x) \right.  \\
&~~~ + 2\int_{-B}^{0} dx\int_{a}^{b} dy\rho_1(x)\rho_2(y)|x-y| +2 \lambda \int_{a}^{b} dx~|x|~\rho_1(x)  \\
&~~~ \left.+ \int_{a}^{b} dx\int_{a}^{b} dy~\rho_2(x)\rho_2(y)|x-y| \right)  \\
&~~~+\mu_2\left(\int_{-B}^{0}dx~\rho_1(x)+\lambda-1+c \right )+\mu_1 \left(\int_{a}^{b}dx~\rho_2(x)-c \right ),
\end{split}
\label{Sigma-2}
\end{align}
where $\mu_1=A_1+A_2$ and $\mu_2=A_1$. Now minimising $\Sigma_c[\rho_I]$ with respect to $\rho_1$ and $\rho_2$, we get the following equations 
\begin{align}
\begin{split}
&\frac{1}{2}x^2 - 2\alpha \int_{-B}^{0} dy\rho_1(y)|x-y| -2 \alpha \lambda |x| -2 \alpha \int_{a}^{b} dy~\rho_2(y)|x-y| + \mu_2=0, \\
 &~~~~~~~~~~~~~~~~~~~~~~~~~~~~~~~
 ~~~~~~~~~~~~~~~~~~~~~~~~~~~~~~~~~~~~~~~~~~~~~~~\text{for}~~-B\le x <0, 
 \end{split}
 \label{eq-rho-1} \\
 \begin{split}
&\frac{1}{2}x^2 - 2\alpha \int_{-B}^{0} dy\rho_1(y)|x-y| -2 \alpha \lambda |x| -2 \alpha \int_{a}^{b} dy~\rho_2(y)|x-y| + \mu_1=0, \\
&~~~~~~~~~~~~~~~~~~~~~~~~~~~~~~~
 ~~~~~~~~~~~~~~~~~~~~~~~~~~~~~~~~~~~~~~~~~~~~~~~\text{for}~~a\le x \leq b\;.
 \end{split} \label{eq-rho-2} 
\end{align}
Taking derivative of the above two equations with respect to $x$ on both sides and using $\frac{d^2}{dx^2}|x-y|=2 \delta(x-y)$, we have 
\begin{eqnarray}
\rho_1(x)&=&\frac{1}{4 \alpha},~~~~\text{for}~~-B\le x <0, \label{rho-1} \\
\rho_2(x)&=&\frac{1}{4 \alpha},~~~~\text{for}~~a\le x \leq b. \label{rho-2}
\end{eqnarray}
We now insert the expression of $\rho_1(x)$and $\rho_2(x)$ into \eqref{eq-rho-1}, to find the following equation 
\begin{equation}
x\left [ \frac{b-a}{2}+2\alpha \lambda -\frac{B}{2}\right] + \left[ \mu_2-\frac{B^2}{4} -  \frac{b^2-a^2}{4} \right] = 0 \;,
\end{equation}
which is valid for all $x$ in the range $-B\le x <0$. As a result we require that the coefficients of $x$ and $x^0$ in the above equation are zero. This implies the following two equations
\begin{eqnarray}
&&2 \alpha \lambda + \frac{(b-a)}{2} - \frac{B}{2}=0,~~~\label{eq-B} \\ 
&& \mu_2 = \frac{B^2}{4}+\frac{b^2-a^2}{4} \;.~~~~~\label{mu_2}
\end{eqnarray}
Similarly, inserting the expressions of $\rho_1(x)$ and $\rho_2(x)$ in  \eqref{eq-rho-2} we find 
\begin{eqnarray}
&&2 \alpha \lambda - \frac{(b+a)}{2} + \frac{B}{2}=0,~~~\label{eq-b} \\ 
&& \mu_1 = \frac{B^2}{4}+\frac{b^2+a^2}{4} \;.~~~~~\label{mu_1}
\end{eqnarray}
We have $6$ unknowns ($a,~b,~B,~\lambda,~\mu_1,~\mu_2$) to determine and till now we have $4$ equations which are 
Eqs. \eqref{eq-B}, \eqref{mu_2}, \eqref{mu_1} and \eqref{eq-b}. We need two additional equations which are obtained from normalisations [see the last line of Eq. (\ref{eq-rho-2})]~: $\int_{-B}^0dx~\rho_1(x)=1-c-\lambda$ and $\int_{a}^bdx~\rho_2(x)=c$. This yields
\begin{eqnarray}
\frac{B}{4\alpha}&=&1-c-\lambda,~~\label{eq-lam}\\
\frac{b-a}{4\alpha}&=&c.~~~\label{eq-ab}
\end{eqnarray}
Solving this system of six equations \eqref{eq-B}, \eqref{mu_2}, \eqref{mu_1}, \eqref{eq-b}, \eqref{eq-lam} and \eqref{eq-ab} for the six unknowns ($a,~b,~B,~\lambda,~\mu_1,~\mu_2$), we get 
\begin{eqnarray}
&&B=2\alpha, \nonumber \\
&&b=2\alpha, \nonumber \\
&&a= 2\alpha(1-2c), \nonumber \\
&&\lambda=\frac{(1-2c)}{2}, \label{constants}\\
&&\mu_1=\alpha^2 (2+(1-2c)^2), \nonumber \\
&&\mu_2=\alpha^2 (2-(1-2c)^2) \;. \nonumber
\end{eqnarray}
With these constants the equilibrium density in \eqref{den-anticipate} is fully specified for $0\leq c \leq 1/2$ 
as [see Eqs. (\ref{den-anticipate}), (\ref{rho-1}), (\ref{rho-2})]
\begin{eqnarray}
\rho_{I}^*(x) = 
\begin{cases}
&\frac{1}{4\alpha} + \frac{(1-2c)}{2}~\delta(x),~~\text{for}~~-2\alpha\leq x \leq 0 \\
&\frac{1}{4\alpha}~~~~~~~~~~~~~~~~~~~~~\text{for}~~~~~~~2\alpha(1-2c)\leq x \leq 2\alpha \;. 
\end{cases},
\label{den-eq}
\end{eqnarray}
Finally inserting this expression of $\rho_{I}^*(x)$ in \eqref{Sigma-2}, we get 
\begin{equation}
\Sigma_c[\rho_{I}^*] = \frac{8 \alpha^2}{3}~(1/2-c)^3 -\frac{2\alpha^2}{3},~~~\text{for}~~0\leq c \leq 1/2. \label{Sigma_c_lt_half}
\end{equation}
A similar computation for $1/2\leq c \leq 1$ yields 
\begin{equation}
\Sigma_c[\rho_{I}^*] = \frac{8 \alpha^2}{3}~(c-1/2)^3 -\frac{2\alpha^2}{3},~~~\text{for}~~1/2\leq c \leq 1. \label{Sigma_c_gt_half}
\end{equation}
Combining both expressions we have 
\begin{equation}
\Sigma_c[\rho_{I}^*] = \frac{8 \alpha^2}{3}~|c-1/2|^3 -\frac{2\alpha^2}{3},~~~\text{for}~~0\leq c \leq 1. \label{Sigma_c}
\end{equation}
Hence from Eqs. (\ref{I_c}) and (\ref{Sigma_c}), we have $I(cN,N) \asymp \e^{-N^3[(8 \alpha^2/3)~|c-1/2|^3 -\frac{2\alpha^2}{3})]}$ to leading order for large $N$. Since $Z_N \approx I(N/2,N)$ and $N! \sim \e^{N \log N + O(N)}$, one finally obtains the large deviation form of the index distribution announced in Eq. (\ref{LDF})
\begin{equation}
P_I(N_+=cN,N)\asymp \exp \left(-N^3\frac{8 \alpha^2}{3}~|c-1/2|^3 \right), \label{LDF}
\end{equation}
It is straightforward to check that this large deviation tail matches smoothly with the tails of  the central region given in \eqref{asymp.1}.

\section{Conclusions}
In this paper we have studied analytically the distribution of the position of the rightmost particle $x_{\max}$ of a $1d$ Coulomb gas confined in an external harmonic potential (the $1d$ jellium model) in the limit of large number of particles $N$. We have obtained the limiting large $N$ distribution describing the typical fluctuations of $x_{\max}$ around its mean, $F_\alpha(x)$ and shown that it is different from the Tracy-Widom distribution of the log-gas. This function $F_\al(x)$ is the solution of a non-local eigenvalue equation (\ref{limit.00}). We have also computed the rate functions associated with atypically large fluctuations around the mean [see Eqs. (\ref{result-1}), (\ref{sm-left_rf-0}) and (\ref{sm-phi+0})] and found a third order phase transition between a pushed and a pulled phase, as in the log-gas. 

In addition, we have studied the distribution of two other observables: (i) the gap $g=x_N-x_{N-1}$ between the two rightmost charges and (ii) the index $N_+$ which is the number of particles on the positive semi-axis. We have analytically computed the distribution of both quantities and found that their typical distributions can be expressed in terms of the same function $F_\al(x)$ [see Eqs. (\ref{scaling_PNg1}) and (\ref{h_a_5}) for $g$ and Eqs. (\ref{limiting.1}) and (\ref{limiting.2}) for $N_+$]. For both observables, the obtained limiting distributions are quite different from their counterpart in the log-gas. In both cases, we have computed the large deviations, to leading order for large $N$ [see Eqs. (\ref{large_dev_gap}) and (\ref{LDF}) for the gap and the index respectively].

Our work raises several interesting questions. For instance, how universal is the limiting distribution of $x_{\max}$ if one changes the confining potential or the pairwise repulsive interaction? It would be challenging to study $x_{\max}$ with a repulsive interaction of the form $|x_i-x_j|^{-k}$ (where $k \to 0$ corresponds to log-gas, while $k=-1$ corresponds to the ``jellium'' model). Unlike the log-gas, the $1d$ jellium does not have a determinantal structure and computing its $n$-point correlations would be interesting. 

\section{Acknowledgements}
The authors would like to acknowledge the support from the Indo-French Centre for the promotion of advanced research (IFCPAR) under Project No. 5604-2. A.~K. would like to acknowledge the financial support from CNRS, France during his visit to LPTMS, Univ. Paris-Sud, Orsay where the the paper has been finalised. This work was partially supported by ANR grant
ANR-17-CE30-0027-01 RaMaTraF.
\appendix


\newpage


\end{document}